\documentclass{aa}  

\usepackage{color}
\usepackage{txfonts}
\usepackage{graphicx}
\usepackage{hyperref}
\usepackage{changes}
\usepackage{soul}
\hypersetup{
    pdftoolbar=true,				% show Acrobat's toolbar? 
    pdfmenubar=true,				% show Acrobat's menu?
    pdffitwindow=false,			% window fit to page when opened
    pdfstartview={FitH},			% fits the width of the page to the window
    pdftitle={title},				% title
    pdfauthor={Author},			% author
    pdfsubject={Subject},			% subject of the document
    pdfcreator={Author},			% creator of the document
    pdfproducer={Author},			% producer of the document
    pdfkeywords={Population \& Evolutionary} ,        % list of keywords
    pdfnewwindow=true,			% links in new window
    colorlinks=true,				% false: boxed links; true: colored links
    linkcolor=cyan,				% color of internal links (change box color with linkbordercolor)
    citecolor=blue,				% color of links to bibliography
    filecolor=cyan,				% color of file links
    urlcolor=cyan				% color of external links
}

\newcommand{\SFH}{\textrm{SFH}}

\newcommand{\CEH}{\textrm{CEH}}
\newcommand{\Starlight}{{\sc Starlight}}
\newcommand{\SL}{{\sc Starlight}}
\newcommand{\Rebetiko}{{\sc Rebetiko}}
\newcommand{\Fado}{{\sc Fado}}

\def\PS{{\tt PS}}
\def\ES{{\tt ES}}
\def\ne{{\tt ne}}

\def\mstar{$M_{\star}$}
\def\baseL{Base $\mathcal{L}$}
\def\baseF{Base $\mathcal{F}$}
\def\baseA{Base $\mathcal{A}$}
\def\ewha{EW(H$\alpha$)}
\def\myoung{$M_{\star}^{<\mathrm{10\,Myr}}/M_{\star}^{\mathrm{total}}$}
\def\mold{$M_{\star}^{>\mathrm{1\,Gyr}}/M_{\star}^{\mathrm{total}}$}

\def\ha{H$\alpha$}
\def\fha{F(H$\alpha$)}

\def\1conf{${\cal S}_1$}
\def\2conf{${\cal S}_2$}
\def\tm{$\langle\log t_{\star}\rangle_{M}$}
\def\tl{$\langle\log t_{\star}\rangle_{L}$}
\def\zm{$\langle Z_{\star}\rangle_{M}$}
\def\zl{$\langle Z_{\star}\rangle_{L}$}
\def\?{\color{red}{\bf ??}\rm\color{black}}
\def\lyc{Ly{\sc c}}

\begin{document} 

\title{Self-consistent population spectral synthesis with {\sc Fado}}
\subtitle{I. The importance of nebular emission in modelling star-forming galaxies}

\author{
Leandro S. M. Cardoso\inst{\ref{adress1}} %,\ref{adress2}}
\and Jean Michel Gomes \inst{\ref{adress1}}
\and Polychronis Papaderos\inst{\ref{adress1}}
}

\institute{
Instituto de Astrof\' isica e Ci\^ encias do Espa\c co, Universidade do Porto, CAUP, Rua das Estrelas, PT4150-762 Porto, Portugal\label{adress1}
}
\date{Received ?? / Accepted ??}

\abstract
% CONTEXT
{Spectral population synthesis (\PS) is a fundamental tool in extragalactic research that aims to decipher the assembly history of galaxies from their spectral energy distribution (SED). Whereas this   technique has led to key insights into galaxy evolution in recent decades, star formation histories (SFHs) inferred therefrom have been plagued by considerable uncertainties stemming from inherent   degeneracies and the fact that until recently all \PS\ codes were   restricted to purely stellar fits, neglecting the essential   contribution of nebular emission (\ne). With the advent of \Fado\   (Fitting Analysis using Differential evolution Optimisation), the now possible self-consistent modelling of stellar and \ne\ opens new   routes to the exploration of galaxy SFHs.}
% AIMS 
{The main goal of this study is to quantitatively explore the accuracy   to which \Fado\ can recover physical and evolutionary properties of   galaxies and compare its output with that from purely stellar \PS\ codes.}
% METHODS
{\Fado\ and \SL\ were applied to synthetic SEDs that track the spectral evolution of stars and gas in extinction-free mock galaxies of solar metallicity that form their stellar mass (\mstar) according to different parametric SFHs. Spectral fits were computed for two different set-ups that approximate the spectral range of SDSS and CALIFA (V500) data, using up to seven libraries of simple stellar   population spectra in the 0.005--2.5 $Z_{\odot}$ metallicity range.}
% RESULTS
{Our analysis indicates that \Fado\ can recover the key physical and evolutionary properties of galaxies, such as \mstar\ and mass- and light-weighted mean age and metallicity, with an accuracy better than 0.2 dex. This is the case even in phases of strongly elevated specific star formation rate (sSFR) and thus with considerable \ne\ contamination ($\mathrm{EW(H\alpha)} > 10^3$ \AA). Likewise, population vectors from \Fado\ adequately recover the mass fraction of stars younger than 10 Myr and older than 1 Gyr (\myoung\ and \mold, respectively) and reproduce with a high fidelity the observed H$\alpha$ luminosity. As for \SL, our analysis documents a moderately good agreement with theoretical values only for evolutionary phases for which \ne\ drops to low levels ($\mathrm{EW(H\alpha)} \leq 60$ \AA) which, depending on the assumed SFH, correspond to an age between $\sim$0.1 Gyr and 2--4 Gyr. However, fits with \SL\ during phases of high sSFR severely overestimate both \mstar\ and the mass-weighted stellar age, whereas strongly underestimate the light-weighted age and metallicity. Furthermore, our analysis suggests a subtle tendency of \SL\ to favour a bi-modal SFH, as well a slightly overestimated \myoung, regardless of galaxy age. Whereas the amplitude of these biases can be reduced, depending on the specifics of the fitting procedure (e.g. accuracy and completeness of flagging emission lines, omission of the Balmer and Paschen jump from the fit), they persist even in the idealised case of a line-free SED comprising only stellar and nebular continuum emission.}
% CONCLUSIONS
{The insights from this study suggest that the neglect of nebular continuum emission in \SL\ and similar purely stellar \PS\ codes could systematically impact \mstar\ and SFH estimates for star-forming galaxies. We argue that these biases can be relevant in the study of a range of topics in extragalactic research, including the redshift-dependent slope of the star formation (SF) main sequence, the SF frosting hypothesis, and the regulatory role of supermassive black holes on the global SFH of galaxies.}

\maketitle
% !!!!!!!!!!!!!!!!!!!!!!!!!!!!!!!!!!!!!!!!!!!!!!!!!!!!!!!!!!!!!!!!!!!!!!!!!!!!!!!!!!!!!!!!!!!!!!!!!!!!!!!!!!!!!!!!!!!!!!!!!!!!!!!!!!!!!!!!!!!!!!!!!!!!!!!!!!!!!!!!!!!!!!!!!!!!!!!!!!!!!!!!!!!!!!!
% - - - - - - - - - - - - - - - - -- - - - - - - - - - - -  - - INTRODUCTION   - - - - - - - - - - - - - - - - - - - - - - - - - - - - - - - - - - - - -
% !!!!!!!!!!!!!!!!!!!!!!!!!!!!!!!!!!!!!!!!!!!!!!!!!!!!!!!!!!!!!!!!!!!!!!!!!!!!!!!!!!!!!!!!!!!!!!!!!!!!!!!!!!!!!!!!!!!!!!!!!!!!!!!!!!!!!!!!!!!!!!!!!!!!!!!!!!!!!!!!!!!!!!!!!!!!!!!!!!!!!!!!!!!!!!!
\section{Introduction}\label{Sec:Introduction}

	Spectral synthesis is a fundamental tool in extragalactic astronomy that aims to recover the main physical and evolutionary characteristics of galaxies across cosmic time \citep[e.g.][]{Panter2003,Heavens2004,Panter2007}.  In particular, its goal is to decipher from the spectral energy distribution (SED) of a galaxy its star formation (SF) and chemical enrichment history (\SFH\ and \CEH, respectively). To this end, two distinct yet complementary techniques have been developed over the past three decades: population and evolutionary synthesis (\PS\ and \ES, respectively). The former attempts to iteratively decompose the SED of a galaxy into its elementary luminosity components, such as individual stars or simple stellar populations (SSPs) selected from a spectral library that spans a relevant range in stellar age and metallicity. The best-fitting combination of the light and mass fractions of these spectral elements, together with the extinction and kinematical broadening they are subjected to, are referred to as the {population vector} (PV) of a galaxy. Quite importantly,  \PS\ is by definition incompatible with any prior assumption on the SFH and CEH of a galaxy \citep[see detailed discussion in the introduction of][hereafter GP17]{GomesPapaderos2017}. Conversely, the ``forward-computing'' concept of \ES\ simplifies the inclusion and consistent treatment of SED ingredients other than stellar emission (e.g. nebular and dust emission) at the price of the \SFH\ and \CEH\ being input assumptions to the model (e.g. \citealt{StruckMarcellTinsley1978,Tinsley1980, Guseva2001, Pacifici12, Conroy2013, Pacifici15, Pacifici16}).

	An extensive application of these spectral synthesis approaches throughout the past decades has lead to significant progress in the understanding of galaxy formation and evolution. For example, several studies on the cosmic evolution of the star formation rate (SFR) density have shown that the SFR density has declined by $\sim$1 dex since a redshift $z \sim 1$--2 \citep{Lilly1996,Madau1996}. Indeed, population synthesis studies  of stellar ``fossil records'' in local galaxies have gone a long way towards corroborating this insight \citep[e.g.][]{Panter2003,Jimenez2005,Panter2007}. Moreover, it was also discovered that the SF at low $z$ takes place predominantly in low-mass galaxies, whereas massive galaxies have experienced the dominant phase of  their mass assembly earlier on, a result commonly known has galaxy downsizing \citep[e.g.][]{Heavens2004}. In addition, the inside-out growth scenario of galaxies has seen support by recent \PS\ studies using integral field spectroscopy (IFS) (e.g. \citealt{Perez2013,Gonzalez2016,Gomes2016c}).

	Spectral synthesis also played a part on revealing the mass-metallicity relation in galaxies. After the pioneering study of the gas-phase metallicity in local star-forming galaxies by \citet{Lequeux1979}, the mass-metallicity relation was further investigated through spectral modelling of SDSS data by \citet{Tremonti2004} and spatially resolved  data from the CALIFA survey (e.g. \citealt{Sanchez2013,Gonzalez2014,Sanchez2017}). Another important step towards the understanding of galaxy formation and evolution has been the analysis of the bimodality of stellar populations in galaxies. This has been extensively studied using  broadband colours \citep[e.g.][]{Strateva2001,Blanton2003,Kauffmann2003a,Baldry2004}, as well as luminosity-weighted stellar ages from spectral modelling and their comparison with the  4000 $\AA$ break (D4000; \citealt{Balogh1999}) and stellar mass (\mstar) of galaxies from volume-limited samples \citep[e.g.][]{Mateus2006,Baldry2006}. More recently, evidence for an age bimodality of galaxy stellar populations on sub-galactic scales was presented in \citet{Zibetti2017}  from spectral modelling of CALIFA IFS data.

	It has also been found a tight correlation between SFR and \mstar\ that was largely established through spectral modelling of star-forming galaxies both in the local universe and higher redshifts \citep[e.g.][]{Brinchmann2004,Noeske2007a,Whitaker2012}. This star-formation main sequence (SFMS) seems to have a behaviour of the form $\mathrm{SFR} \propto M_{\star}^{\alpha(z)}$ with a likely redshift-dependent exponent of $\alpha(z) \neq 1$. Additionally, spectral modelling of IFS data points to a local SFMS relation (e.g. \citealt{Cano2016,Hsieh2017}). Moreover, several studies have presented a connection between the SFH and gas-phase metallicity for star-forming  galaxies by showing that, whereas all star-forming galaxies assembled the bulk of their stellar mass more than 1 Gyr ago, low-nebular metallicity systems evolve at a slower pace and are currently forming stars at a significantly higher specific SFR (sSFR)  as compared to metal-rich galaxies (e.g. \citealt{Asari2007,LaraLopez2010}).

	Spectral synthesis has also helped to shed light on the long-standing debate regarding the gas excitation mechanisms in emission-line galaxies. Classical emission-line diagnostics \citep[e.g.][hereafter BPT-VO]{Baldwin1981,Veilleux1987}, which have been extensively used for distinguishing photoionisation by stars and active galactic nuclei (AGN), are prone in the case of weak-line emitters to substantial uncertainties due to the correction for underlying stellar absorption \citep[e.g.][]{Stasinska2008,CidFernandes2010,CidFernandes2011,Petropoulou2011}.  Only with the advent of spectral synthesis has it been possible to overcome this issue and carry  out unsupervised BPT-VO studies of large galaxy samples (e.g. \citealt{Kewley2001,Kauffmann2003c,Stasinska2006,Schawinski2007}).

	Studies of the warm interstellar medium (ISM) in early-type galaxies (ETGs) have also benefit from developments in spectral synthesis. Indeed, the detection of faint \ne\  in ETGs has been possible through spectral modelling and subtraction of the stellar SED \citep[e.g.][]{Sarzi2006}, allowing for the detection of diffuse ionised gas at an equivalent width (EW) level of $\sim$0.5--3.0 \AA\ in several local ETGs \citep[e.g.][]{Heckmann1980,Phillips1986,Binette1994,Stasinska2008,Sarzi2010,Annibali2010,Kehrig2012,CidFernandes2010,CidFernandes2011,YanBlanton2012,Papaderos2013,Gomes2016a,Gomes2016b,Gomes2016c,Lacerda2018}. In particular, spatially resolved analysis of this faint substrate of \ne\  with IFS data has allowed for the determination of \ha\ intensity and EW maps which, together with BPT-VO diagnostics, have placed tight constraints on the role of photoionisation by the post-AGB stellar component relative to that by AGN, shocks, and SF \citep[e.g.][]{Kehrig2012,Papaderos2013,Singh2013,Gomes2016a,Gomes2016b,Gomes2016c}.

	Despite the progress achieved thanks to constantly improving \PS\ codes (see the discussion in GP17) it should be kept in mind  that until recently these suffered from significant deficiencies, in particular, the well-known SFH-metallicity-extinction-kinematics degeneracy issue \citep[e.g.][]{OConnell1996,Pelat1997,Pelat1998} and the neglect of  \ne\  in spectral fits.  As proposed in GP17, the latter caveat might be partly responsible for the former. Indeed, a specially important limitation is the neglect of \ne\ (continuum plus lines), which is known to be an important component of a galaxy SED that can comprise up to 30--50\% of the total optical and near-infrared emission of starburst (i.e. high-sSFR) galaxies \citep[e.g.][]{Grewing1968,Huchra1977,Krueger1995,Leitherer1995,Izotov97, Papaderos1998,Leitherer1999,Schaerer2009,Schaerer2010}. Taking \ne\ into account in \PS\ fits is therefore  fundamental to a realistic SED modelling and accurate recovery of SFH and CEH for these systems. 

	Notwithstanding this fact, the importance of \ne\ has been seldom appreciated in previous testsuits of \PS\ codes,  all of which were validated against purely stellar synthetic SEDs  \citep[e.g.][]{Gomes2005,CidFernandes2005,Gomes2009,CidFernandes2014,Magris2015,Lopez2016,Wilkinson2017}, with a few exceptions such as  STECKMAP \citep{Ocvirk2006a,Ocvirk2006b} and UlySS \citep{Koleva2009}, which were also tested on realistic (stellar plus nebular) synthetic SEDs.  However, these two stellar fitting codes might be regarded as a special case, since their mathematical concept deviates from that of \PS\ through an implicit regularisation of their best-fitting PV. This provision, employing in the case of STECKMAP a Laplacian convolution kernel, forces the physical and evolutionary ingredients of the best-fitting PV (e.g. SFH) to take a smooth shape, which has the advantage of suppressing an ``exaggeratedly oscillating solution'' \citep{Ocvirk2006b,Ocvirk2010}. Whereas this might be advantageous for spectral synthesis studies of galaxies assembling at a smooth pace, it has the caveat of erasing short-duration SF events from the PV, which is a disadvantage in the case of modelling starburst and high-sSFR galaxies. Owing to the a priori exclusion of certain forms of a PV, these two codes may be considered as hybrid realisations of the \PS\ and \ES\ spectral fitting concept.

	From this, it might be argued that most of the benchmarking results obtained so far yield lower limits  to the true uncertainties of \PS\ fitting and, in particular, they do not capture systematic biases in stellar mass, age, and metallicity determinations that are predicted to  arise from the neglect of \ne\ (see GP17 for a brief overview on previous work on this subject and e.g. \citealt{Papaderos2002,Izotov2011}). Further uncertainties in spectral fits with standard purely stellar \PS\ codes are expected in the  case of a mixture of stellar emission with an AGN power law \citep{Cardoso2016,Cardoso2017} and also whenever  other determinants of a galaxy SED (e.g. dust absorption and emission, shocks, radiation transfer effects)  are not or only partly taken into account in spectral fits. In view of such considerations, it appears worthwhile to devise a new set of testsuits for evaluating random and systematic uncertainties in \PS\ by fitting progressively realistic synthetic SEDs for a variety of SFHs and CEHs.

	This is the first of a series of articles that aim to define and apply a set of new benchmark tests of the overall performance of \PS\ codes and their ability  to recover physical and evolutionary properties of galaxies, similar to a suite of standard cases defined for  photoionisation codes\footnote{As defined by the workshops on photoionisation at Meudon (\citealt{Pequignot1986}) and Lexington (\citealt{Ferland1995,Pequignot2001}).}. In this paper, we test two conceptually distinct \PS\ codes  on synthetic SEDs that consistently include stellar and \ne\  in a dust-free environment. Our main goal is to explore the impact of nebular continuum emission on the best-fitting PVs  for a simple experimental framework that involves a narrow set of parametric SFHs between the limiting case of  an instantaneous burst and a continuous SF at a constant SFR. Future articles of this series will explore the effect that a variation in the extinction, metallicity, and velocity dispersion of the input SEDs may have on the best-fitting SFH and CEH. In this article, we make use of \SL\ \citep[][]{CidFernandes2005}, as representative of purely stellar \PS\ codes, and \Fado\ \citep[Fitting Analysis using Differential evolution Optimisation;][]{GomesPapaderos2017,GomesPapaderos2018}, the only publicly available \PS\ code that includes \ne\ and ensures  consistency between the best-fitting PV and observed \ne\ characteristics in a star-forming galaxy.

	This paper is organised as follows. Section \ref{Sec:Methodology} details the methodology adopted for the computation of synthetic SEDs and their modelling with \Fado\ and \SL. Moreover, Sect.~\ref{Sec:Results} presents the main results of this study, which are discussed and summarised in Sects. \ref{Sec:Discussion} and \ref{Sec:Conclusions}, respectively.

% !!!!!!!!!!!!!!!!!!!!!!!!!!!!!!!!!!!!!!!!!!!!!!!!!!!!!!!!!!!!!!!!!!!!!!!!!!!!!!!!!!!!!!!!!!!!!!!!!!!!!!!!!!!!!!!!!!!!!!!!!!!!!!!!!!!!!!!!!!!!!!!!!!!!!!!!!!!!!!!!!!!!!!!!!!!!!!!!!!!!!!!!!!!!!!!
% - - - - - - - - - - - - - - - - -- - - - - - - - - - - -  - - - -  METHODOLOGY   - - - - - - - - - - - - - - - - - - - - - - - - - - - - - - - - - - - - - - - - 
% !!!!!!!!!!!!!!!!!!!!!!!!!!!!!!!!!!!!!!!!!!!!!!!!!!!!!!!!!!!!!!!!!!!!!!!!!!!!!!!!!!!!!!!!!!!!!!!!!!!!!!!!!!!!!!!!!!!!!!!!!!!!!!!!!!!!!!!!!!!!!!!!!!!!!!!!!!!!!!!!!!!!!!!!!!!!!!!!!!!!!!!!!!!!!!!
\section{Methodology}\label{Sec:Methodology}

	This section details the methodology adopted for the computation of synthetic galaxy SEDs with the \ES\ code \Rebetiko\ and their subsequent modelling with the \PS\ codes \Fado\ and \SL\ in order to estimate the most important properties of the stellar component.

% !!!!!!!!!!!!!!!!!!!!!!!!!!!!!!!!!!!!!!!!!!!!!!!!!!!!!!!!!!!!!!!!!!!!!!!!!!!!!!!!!!!!!!!!!!!!!!!!!!!!!!!!!!!!!!!!!!!!!!!!!!!!!!!!!!!!!!!!!!!!!!!!!!!!!!!!!!!!!!!!!!!!!!!!!!!!!!!!!!!!!!!!!!!!!!!
\subsection{Construction of a library of synthetic SEDs}\label{Subsec:Synthetic_Spectra}

	The SED of composite stellar populations (CSPs) with \ne\  (continuum and most prominent optical lines) were computed with the \ES\ code \Rebetiko\ following:
        
	\begin{equation}\label{Eq:CSP_REBETIKO}
	\begin{split}
		F_\lambda(t,Z)=	& \int_{0}^{t} \Psi(t-t^{\prime}) ~ F^{\text{SSP}}_\lambda(t^{\prime},Z) ~ \textrm{d} t^{\prime}  \\
						& + \frac{\gamma_{\text{eff}}(T_e)}{\alpha_{\text{B}}(T_e)}  ~ \int_{0}^{t} \Psi(t-t^{\prime}) ~ 
						q^{\text{SSP}}(t^{\prime},Z) ~ \text{d}t^{\prime} \, .
                                                                                                                                                                																				\end{split} 																							\end{equation}
        
	The first integral in Eq. \ref{Eq:CSP_REBETIKO} represents the sum of the SSP spectra $F^{\text{SSP}}_\lambda$ of age $t$ and metallicity $Z$ weighted by the SFR $\Psi(t) = \text{d}M_\star/\text{d}t$. The second integral represents the nebular continuum determined by multiplying the number of hydrogen ionising photons ($\lambda \leq 911.76$ \AA\ or $E\geq13.6$ eV), represented by the term $\int_0^{t} q^{\text{SSP}}(t^{\prime}, Z) ~ \text{d}t^{\prime}$, by the SFR and the ratio of the effective continuous emission coefficient $\gamma_{\text{eff}}$ to the case B recombination coefficient  $\alpha_{\text{B}}$. Overall, the nebular modelling adopted in this work is similar to \cite{Schaerer2009}. The nebular continuum is computed assuming case B recombination under typical physical conditions of HII regions, those being, an electron density of $n_e = 100$ cm$^{-3}$ and temperature of $T_e = 10^4$ K (\citealt{Osterbrock1989}; \citealt{Osterbrock2006}). Moreover, H emission-lines fluxes are computed considering different electron temperatures and densities (\citealt{Hummer1987}) and the UV ionising output from the stellar populations (\citealt{Zanstra1961}) for radiation-bounded HII regions. The determination of line fluxes for heavier elements follow the semi-empirical calibration from \cite{AF03} as a function of the nebular metallicity normalised to H$\beta$. 

\begin{figure*}
\begin{center}
\includegraphics[width=0.99\textwidth]{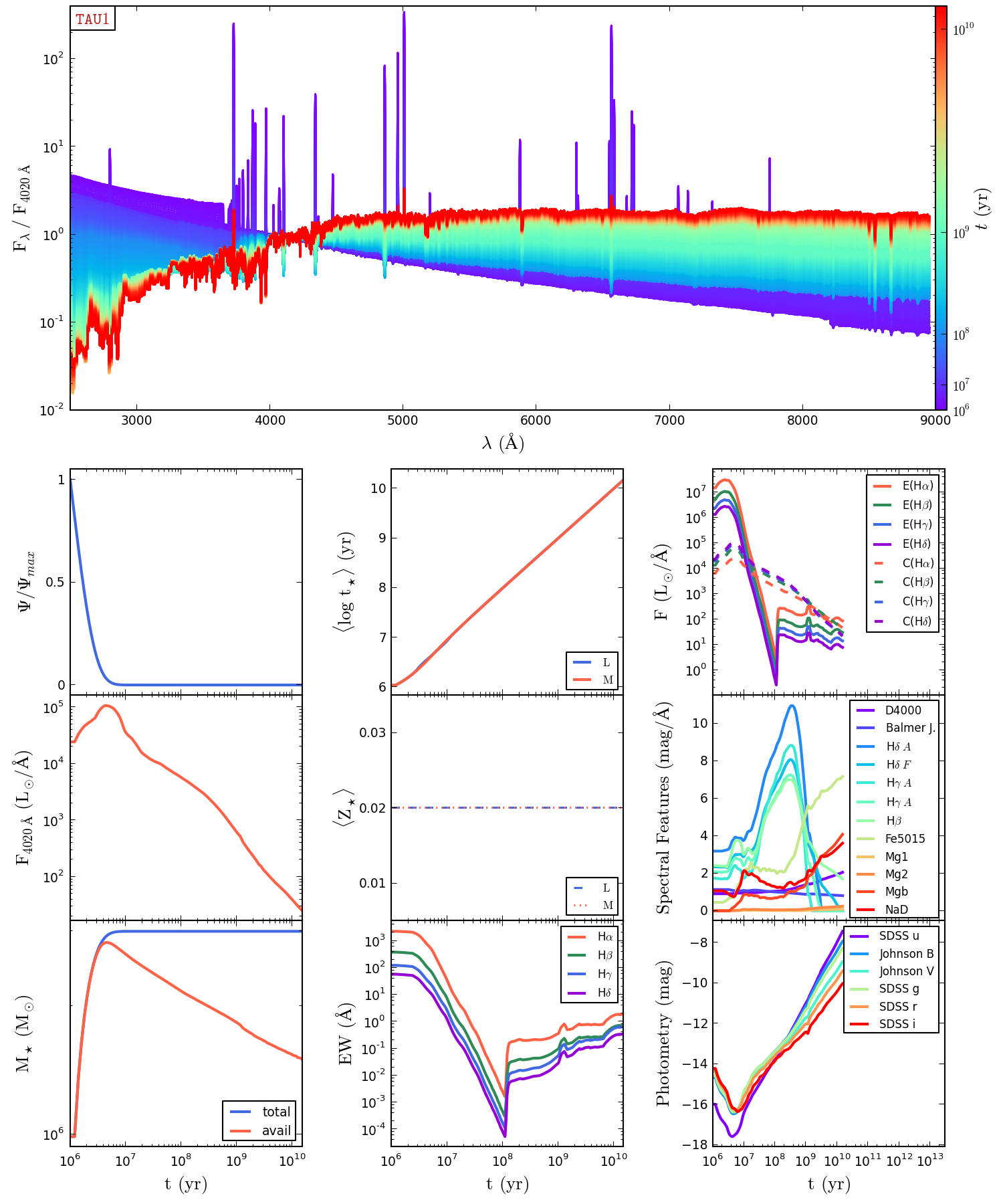}
\caption{Synthetic CSP spectra with \ne\ normalised at $\lambda_0=4020$ \AA\ and other spectrophotometric quantities computed with the \ES\ code \Rebetiko\ for an instantaneous burst SFH ({\tt TAU1}) with the colour coding on the main panel representing the CSP age $t$. Smaller panels from {left-} to {right-hand side} and {top} to {bottom} display as a function of $t$, respectively:  SFR $\Psi(t)$ ({blue line});  mean stellar age $\langle\log t_{\star}\rangle$ weighted by light ({blue line}) and mass ({red line});  predicted total luminosity ({full lines}) and monochromatic luminosity at the continuum (dashed lines) of the Balmer emission lines H$\alpha$ ({red lines}), H$\beta$ ({blue lines}), H$\gamma$ ({green lines}) and H$\delta$ ({violet lines});  flux at normalisation wavelength $F_{\lambda_0}$ ({red line}); stellar metallicity $\langle Z_{\star}\rangle$ weighted by light ({blue line}) and mass ({dashed red line}); selected Lick/IDS indices; total ever formed ({blue line}) and currently available ({red line}) stellar mass $M_{\star}$;  predicted EWs of the Balmer emission lines H$\alpha$ ({red line}), H$\beta$ ({blue line}), H$\gamma$ ({green line}) and H$\delta$ ({violet line}); and photometric magnitudes for passbands of selected photometric systems. }
\label{Fig:RBTK_Spectra_-_Tau1}
\end{center}
\end{figure*}

\begin{figure*}[h!]
\begin{center}
\includegraphics[width=0.99\textwidth]{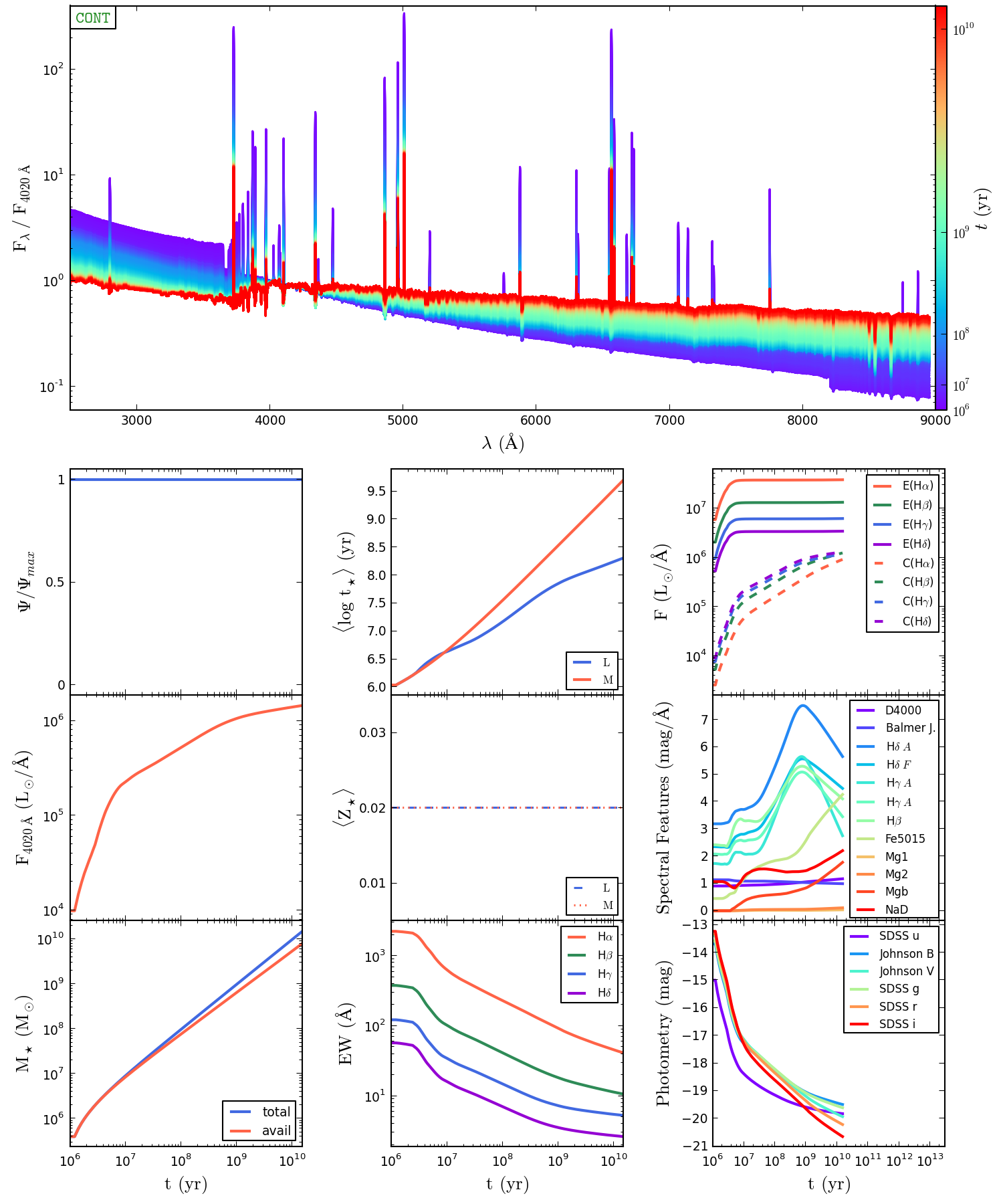}
\caption{Synthetic CSP spectra with \ne\ normalised at $\lambda_0=4020$ \AA\ and other spectrophotometric quantities computed with \Rebetiko\ for a continuous SFH. The layout and legend details are identical to those of Fig. \ref{Fig:RBTK_Spectra_-_Tau1}. }
\label{Fig:RBTK_Spectra_-_Cont}
\end{center}
\end{figure*}

	Figures \ref{Fig:RBTK_Spectra_-_Tau1} and \ref{Fig:RBTK_Spectra_-_Cont} show synthetic SEDs normalised at $\lambda_0=4020$ \AA\ for ages between 1 Myr and 15 Gyr for instantaneous and continuous SFHs, respectively. A set of 716 CSPs was created for each of the 9 SFR functions presented in \cite{Cardoso2017}; 3 exponentially declining, 1 continuous and 5 delayed. The spectra cover the 91 to 9000 \AA\ range with $\Delta\lambda=1$ \AA\ and were computed by adopting \cite{BruzualCharlot2003} solar-metallicity ($Z_{\odot}=0.02$) SSPs with a \cite{Chabrier2003} IMF and  Padova 1994 evolutionary tracks (\citealt{Alongi1993, Bressan1993, Fagotto1994a, Fagotto1994b, Girardi1996}). Moreover, the models assume no extinction ($V$-band extinction of $A_V = 0$ mag) and the line broadening due to stellar velocity dispersion is kept to that inherent to the \cite{BruzualCharlot2003} SSPs.

% FIGURE  1 - - - - - - - - - - - - - - - - - - - - - - - - - - - - - - - - - - - - - - - - - - - - - - - - - - - - - - - - - - - - - - - - - - - - - - - - - - - - - - - - 
% - - - - - - - - - - - - - - - - - - - - - - - - - - - - - - - - - - - - - - - - - - - - - - - - - - - - - - - - - - - - - - - - - - - - - - - - - - - - - - - - - - - - - - - - -
% FIGURE  2 - - - - - - - - - - - - - - - - - - - - - - - - - - - - - - - - - - - - - - - - - - - - - - - - - - - - - - - - - - - - - - - - - - - - - - - - - - - - - - - - 
% - - - - - - - - - - - - - - - - - - - - - - - - - - - - - - - - - - - - - - - - - - - - - - - - - - - - - - - - - - - - - - - - - - - - - - - - - - - - - - - - - - - - - - - - -

	The stellar properties of a galaxy can be assessed through various model output quantities, such as the current SFR, SFH, CEH and/or mass determinations. For instance, the SFH and CEH of a galaxy can be represented to a first order by its mean stellar age and mean stellar metallicity (e.g. \citealt{CidFernandes2005}), which can be determined considering the ages/metallicities and fractional contributions weighted by light and mass of the SSPs that compose each synthetic spectrum. For instance, the mean logarithmic stellar age weighted by light and mass can be written as, respectively,

	\begin{equation}\label{Eq:mean_stellar_age_by_light}	
		\langle\log \, t_{\star}\rangle_L = \sum^{N_{\star}}_{i=1}   \gamma_i\,. \log t_i \, ,
																				\end{equation}	
	
	\begin{equation}\label{Eq:mean_stellar_age_by_mass}					
		\langle\log \, t_{\star}\rangle_M = \sum^{N_{\star}}_{i=1}   \mu_i\,. \log t_i \, ,
																				\end{equation}	

\noindent where $t_i$ is the age of the i$^{\text{th}}$ SSP element and $\gamma_i$ and $\mu_i$ are its light and mass fractions, respectively. The term $\mu_i$ refers specifically to the \mstar\ corrected for the stellar mass fraction returned to the  ISM during the evolution of the galaxy. Moreover, if $Z_i$ is the metallicity of the i$^{\text{th}}$ SSP element, then the logarithmic mean stellar metallicity weighted by both light and mass can be written as, respectively,   
	
	\begin{equation}\label{Eq:mean_stellar_metallicity_by_light}			
		\log\langle Z_{\star}\rangle_L = \log \sum^{N_{\star}}_{i=1}   \gamma_i\,.\, Z_i \, ,
																				\end{equation}	
	\begin{equation}\label{Eq:mean_stellar_metallicity_by_lmass}				
		\log\langle Z_{\star}\rangle_M = \log \sum^{N_{\star}}_{i=1}   \mu_i\,.\, Z_i \, .
																				\end{equation}

% !!!!!!!!!!!!!!!!!!!!!!!!!!!!!!!!!!!!!!!!!!!!!!!!!!!!!!!!!!!!!!!!!!!!!!!!!!!!!!!!!!!!!!!!!!!!!!!!!!!!!!!!!!!!!!!!!!!!!!!!!!!!!!!!!!!!!!!!!!!!!!!!!!!!!!!!!!!!!!!!!!!!!!!!!!!!!!!!!!!!!!!!!!!!!!!
\subsection{Fitting of synthetic SEDs}\label{Subsec:PS_Application}

	The \PS\ codes \Starlight\footnote{Version 04: \url{http://www.starlight.ufsc.br}} and \Fado\footnote{Version 1b: \url{http://www.spectralsynthesis.org}} were applied to these synthetic spectra with the goal of quantifying the impact of \ne\ on the estimation of the physical and evolutionary properties of the stellar component.           \cite{CidFernandes2005} showed that \Starlight\ can recover the stellar mass within $\sim$0.1 dex and the mean age and metallicity within $\sim$0.2 dex from purely stellar SEDs. These values may be regarded as typical of the accuracy of purely stellar \PS\ codes (\citealt{Ocvirk2006a, Ocvirk2006b, Tojeiro2007, Koleva2009, Wilkinson2017}). Recently, \cite{GomesPapaderos2017} have shown in a pilot analysis that \Fado\ yields a similar or better precision, even when applied to more realistic galaxy spectra including \ne, thanks to its ability to self-consistently model stellar and \ne\  (i.e. continuum and hydrogen Balmer lines).

	The synthetic SEDs were fitted in this study with both \Fado\ and \Starlight\ while masking the strongest emission lines.  On the one hand, this was accomplished with \SL\ by adopting the spectral mask provided by the SEAGal collaboration, which is based on the analysis of SDSS galaxies (\citealt{CidFernandes2005}). It is important to note that weaker emission lines not considered in this mask can be very prominent during phases of elevated sSFR and, if incompletely removed, can have in principle a significant impact on the spectral fits. For this reason, both \Fado\ and \SL\ integrate by default  the provision for automatic identification and 3$\sigma$ clipping of emission lines. On the other hand, spectral modelling with \Fado\ was carried out without prior application of an emission-line mask and allowing for an unsupervised rejection of emission lines in the code. 

	Models were computed within two spectral intervals:  3400--8900 \AA\ (\1conf) and 3800--7600 \AA\ (\2conf). Configuration \1conf\ includes the Balmer and Paschen jump, which are pronounced for young ($< 10^7$ yr) ages and cannot be properly fitted with purely stellar templates. Fits within the narrow spectral range of the set-up \2conf\ do not include these nebular continuum discontinuities, thereby in principle easing purely stellar fits. This is of special importance since, as shown by, for instance, \cite{Reines2010}, the nebular continuum around the Paschen jump can account for more than 40\% of the $I$-band flux   in young massive stellar clusters. Set-ups \1conf\ and \2conf\ may be considered representative of spectral fits of SDSS (\citealt{York2000}) spectra for galaxies $z\ga 0.05$ and CALIFA V500 IFS data (\citealt{Sanchez2012}), respectively.  Moreover, the fitting was performed with both codes  by keeping the recessional velocity and the line-of-sight velocity dispersion as free parameters within $v_{\star}=\textnormal{-}500\textnormal{--}500$ km/s and $\sigma_\star=0\textnormal{--}500$ km/s, respectively.  The intrinsic $V$-band extinction was kept fixed at $A_V=0$ mag. However, it is interesting to note that test fits with \Fado\ in which $A_V$ was allowed to vary between -0.5 and 6 mag showed that the code recovers the input value of 0 mag within a narrow 1$\sigma$ deviation of $\sim$0.026 mag. 
       	
\begin{figure*}
\includegraphics[width=1\textwidth]{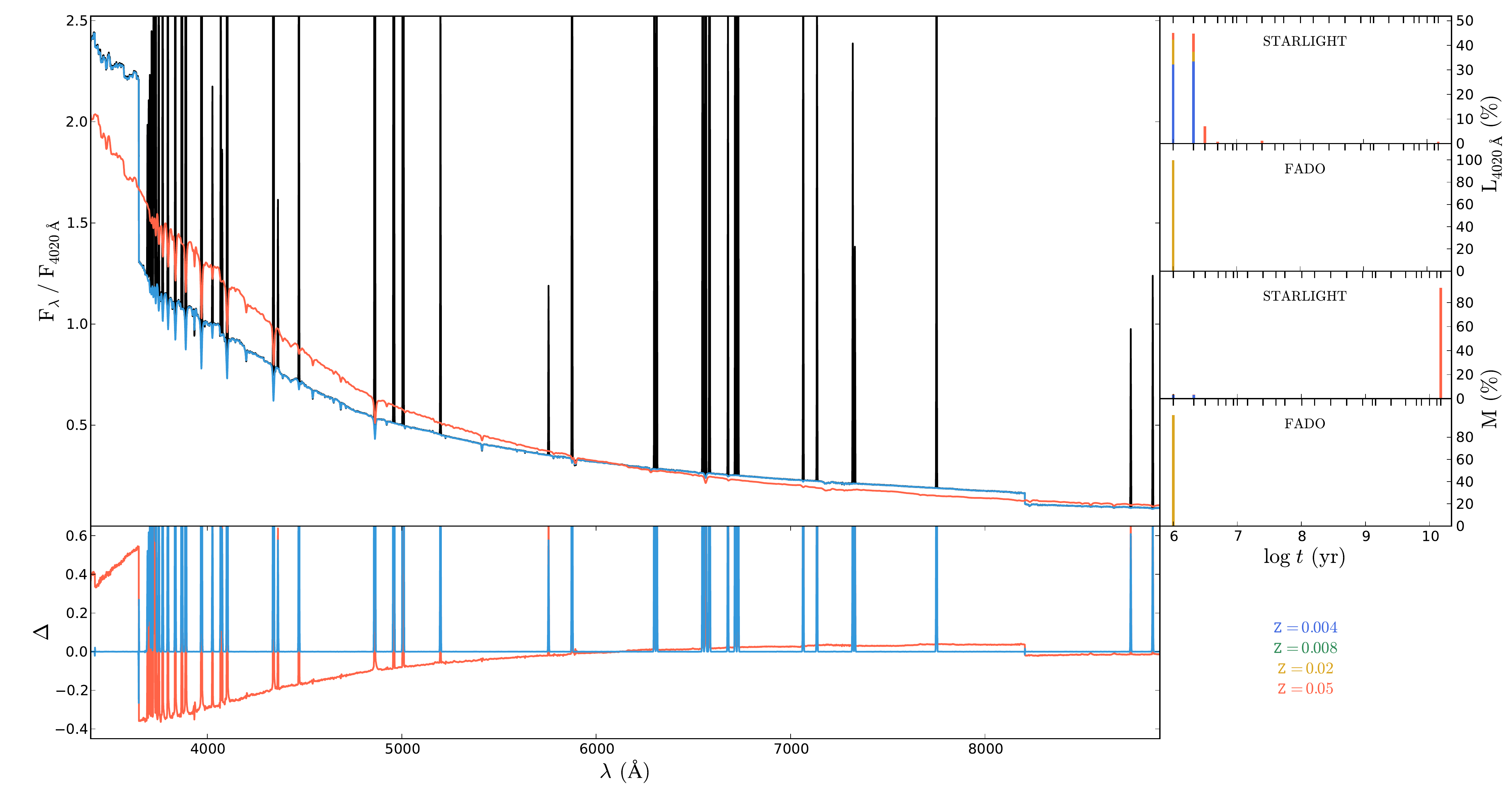}
\caption{Spectral modelling results with \SL\ and \Fado\ for a synthetic SED that includes \ne\ for an instantaneously formed stellar population of solar metallicity at an age of 1 Myr. Main panel: Black, red, and blue lines show, respectively, the input SED and its fit with \SL\ and \Fado. Bottom panel: Red and blue lines represent the residuals that result from subtraction of the \SL\ and \Fado\ fits from the input SED. Top right-hand side panels: Light fractions at the normalisation wavelength as a function of age for SSPs selected by \SL\ (top) and \Fado\ (bottom) are shown with colour coding representing metallicity. Bottom right-hand side panels: Corresponding mass fractions of the SSPs included in the best-fitting PVs with \SL\ (top) and \Fado\ (bottom) are shown.}
\label{Fig:Starlight_and_Fado_fit}
\end{figure*}

	These tests were carried out using several SSP base libraries with 45 to 1326 elements, details of which can be found in Appendix \ref{Appendix:Bases}. For the sake of simplicity, the following analysis is focussed on results for an instantaneous burst and continuous SFHs with a base with 100 SSPs (\baseL), which is identical to that adopted in \cite{Cardoso2017}. However, results for other 6  base libraries and for all of the 9 adopted SFHs can be found in the Appendices \ref{Appendix:Bases} and \ref{Appendix:SFHs}, respectively.  \baseL\ is composed of SSPs with 25 ages between 1 Myr and 15 Gyr (black vertical lines on the right-hand side panels of Fig. \ref{Fig:Starlight_and_Fado_fit}) for 4 metallicities ($Z = 0.004$, 0.008, 0.02, and 0.05) with a \cite{Chabrier2003} IMF and Padova 1994 evolutionary tracks (\citealt{Alongi1993, Bressan1993, Fagotto1994a, Fagotto1994b, Girardi1996}).

	Figure \ref{Fig:Starlight_and_Fado_fit} shows the  \Starlight\ and \Fado\ spectral fits in set-up \1conf\ for a CSP with 1 Myr and instantaneous burst SFH. Black, red, and blue lines on the main panel represent the input, \Starlight\ and \Fado\ best-fit spectra, respectively. Moreover, the right-hand side panels illustrate the best-fitting SFH,  delineated in the top panel by the luminosity contribution $L_{4020}$ of the selected SSPs at the normalisation wavelength (4020 \AA) and in the bottom panel by their corresponding contribution to the total stellar mass \mstar. This illustrative case shows that \Starlight\ estimates $\log M_{\star}=7.23$ $M_{\odot}$,  $\langle \log t_{\star} \rangle \simeq 7.21$ yr and $\log \langle Z_{\star} \rangle\simeq-1.68$, which differ significantly from the input of $\log M_{\star}= 5.73$ $M_{\odot}$, $\langle \log t_{\star} \rangle=6$ yr  and $\log \langle Z_{\star} \rangle=-1.70$. Meanwhile, \Fado\ yields $\log M_{\star}= 5.74$ $M_{\odot}$, $\langle \log t_{\star} \rangle=6$ yr  and $\log \langle Z_{\star} \rangle=-1.70$, in agreement with the \Rebetiko\ values.  The results from \Fado\ in the fitting set-ups \1conf\ and \2conf do not noticeably differ from one another, whereas  models with \SL\ depend as expected on whether the strong Balmer jump is included in the fitted spectral range.

% FIGURE  3 - - - - - - - - - - - - - - - - - - - - - - - - - - - - - - - - - - - - - - - - - - - - - - - - - - - - - - - - - - - - - - - - - - - - - - - - - - - - - - - - 
% - - - - - - - - - - - - - - - - - - - - - - - - - - - - - - - - - - - - - - - - - - - - - - - - - - - - - - - - - - - - - - - - - - - - - - - - - - - - - - - - - - - - - - - - -
		
% !!!!!!!!!!!!!!!!!!!!!!!!!!!!!!!!!!!!!!!!!!!!!!!!!!!!!!!!!!!!!!!!!!!!!!!!!!!!!!!!!!!!!!!!!!!!!!!!!!!!!!!!!!!!!!!!!!!!!!!!!!!!!!!!!!!!!!!!!!!!!!!!!!!!!!!!!!!!!!!!!!!!!!!!!!!!!!!!!!!!!!!!!!!!!!!
% - - - - - - - - - - - - - - - - -- - - - - - - - - - - -  - - - -  RESULTS   - - - - - - - - - - - - - - - - - - - - - - - - - - - - - - - - - - - - - - - - 
% !!!!!!!!!!!!!!!!!!!!!!!!!!!!!!!!!!!!!!!!!!!!!!!!!!!!!!!!!!!!!!!!!!!!!!!!!!!!!!!!!!!!!!!!!!!!!!!!!!!!!!!!!!!!!!!!!!!!!!!!!!!!!!!!!!!!!!!!!!!!!!!!!!!!!!!!!!!!!!!!!!!!!!!!!!!!!!!!!!!!!!!!!!!!!!!
\section{Results}\label{Sec:Results}

% !!!!!!!!!!!!!!!!!!!!!!!!!!!!!!!!!!!!!!!!!!!!!!!!!!!!!!!!!!!!!!!!!!!!!!!!!!!!!!!!!!!!!!!!!!!!!!!!!!!!!!!!!!!!!!!!!!!!!!!!!!!!!!!!!!!!!!!!!!!!!!!!!!!!!!!!!!!!!!!!!!!!!!!!!!!!!!!!!!!!!!!!!!!!!!!
\subsection{Physical and evolutionary properties inferred from spectral fitting}

% FIGURE  4 - - - - - - - - - - - - - - - - - - - - - - - - - - - - - - - - - - - - - - - - - - - - - - - - - - - - - - - - - - - - - - - - - - - - - - - - - - - - - - - - 
\begin{figure*}
\begin{center}
\includegraphics[width=0.99\textwidth]{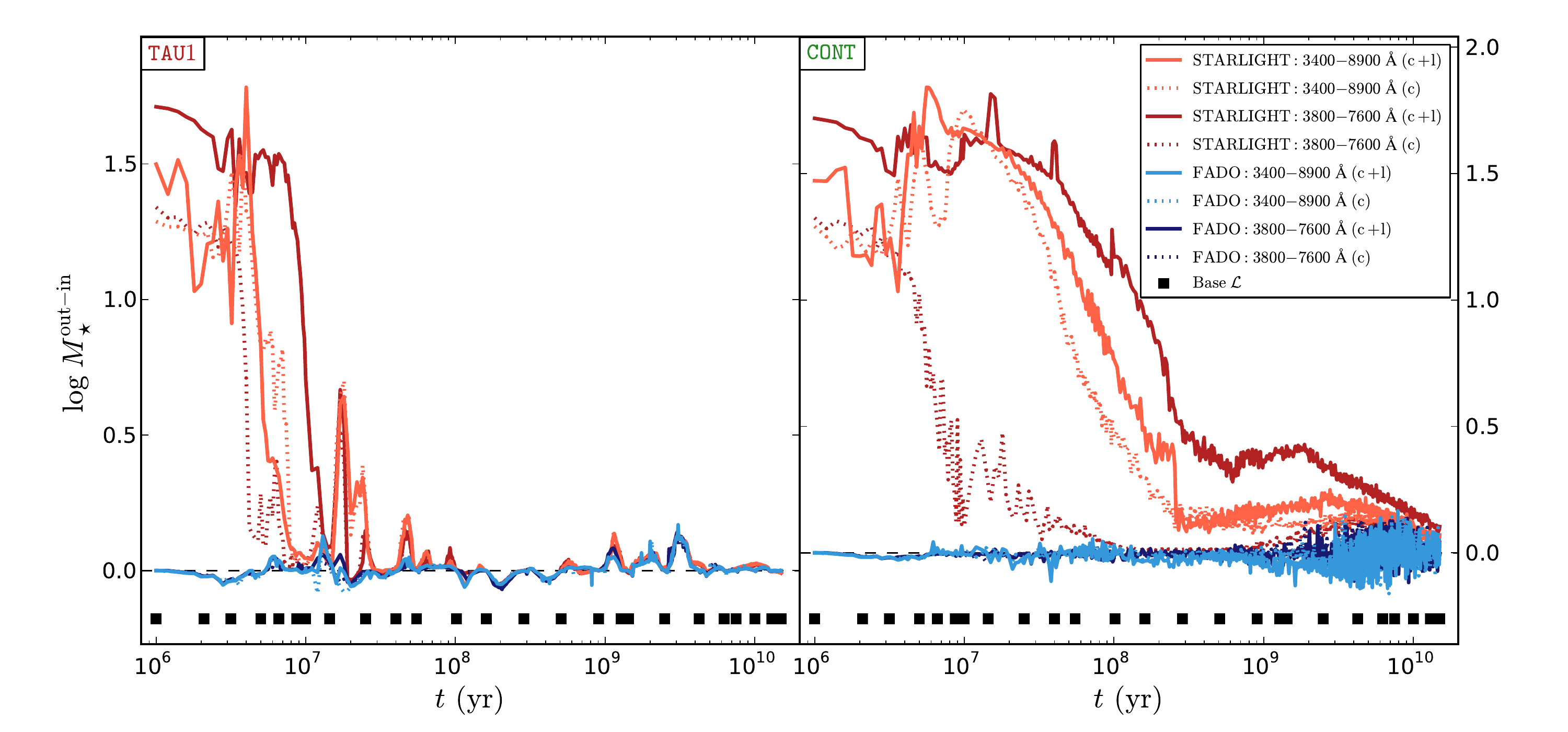}
\caption{Difference in the total stellar mass $M_{\star}$ inferred from fits with \SL/\Fado\ ({out}) and the actual value of the synthetic SEDs from \Rebetiko\ ({in}) as a function of the CSP age $t$. Results for the case of an instantaneous burst ($\texttt{TAU1}$) and continuous SFH ($\texttt{CONT}$) are shown in the left and right panel, respectively. Red and blue lines represent results with \Starlight\ and \Fado, respectively. Light and dark red/blue lines illustrate results from fits in the spectral ranges 3400--8900 \AA\ (\1conf) and 3800--7600 \AA\ (\2conf), respectively. Dotted lines with the same colour coding illustrate results from spectra with stellar and only nebular continuum. Black squares represent the ages covered by the SSP library \baseL .}
\label{Fig:COVER_-_Mstar_diff_vs_age}
\end{center}
\end{figure*}
% - - - - - - - - - - - - - - - - - - - - - - - - - - - - - - - - - - - - - - - - - - - - - - - - - - - - - - - - - - - - - - - - - - - - - - - - - - - - - - - - - - - - - - - - -

	Figure \ref{Fig:COVER_-_Mstar_diff_vs_age} shows the logarithmic ratio of the currently available total stellar mass \mstar\ computed from spectral fits to that of the input SEDs as a function of the CSP age $t$. Results for the \1conf\ and \2conf\ are represented, respectively, for \Fado\ in light and dark blue and for \SL\ in light and dark red.  Moreover, left- and right-hand side panels represent results for instantaneous burst ({\tt TAU1}) and continuous ({\tt CONT}) SFHs. It can be seen that spectral fits with \Fado\ in both the \1conf\ and \2conf\ configurations recover the input \mstar\ within 0.15 dex over the entire age interval considered. Moreover, the slight hump ($\sim$0.14 dex) in determinations pertaining to {\tt CONT} models for ages $\ga$3 Gyr is presumably due to the sparse age coverage of the SSP library represented by the black squares. As for \mstar\ determinations with \SL, they are overestimated for \1conf\ by up to $\sim$1.8 dex both for {\tt TAU1} and {\tt CONT} for a young age of $t \la 10$ Myr and 50 Myr, respectively. Whereas in the case of {\tt TAU1} (left) \SL\ yields a reasonably good match within $\sim$0.15 dex to the input \mstar\ for ages $t\ga100$ Myr, the $\texttt{CONT}$ models reveal an excess by up to 50\% in the estimated \mstar\ for $t \geq 300$ Myr.  As apparent from Fig. \ref{Fig:SFHs_-_Mstar_diff_vs_age}, the \ewha\ at $t\geq 3$ Gyr for {\tt CONT} is $\sim$40--60 \AA, which is typical for the discs of local late-type galaxies \citep[cf. e.g.][]{BredaPapaderos2018}. Indeed, the sudden increase of \ewha\ at an age $\sim$10$^8$ yr illustrated in this figure is due to photoionisation by post-AGN stars (e.g. \citealt{CidFernandes2011,Gomes2016a}). A similar yet slightly stronger bias is apparent from models in the \2conf\ configuration, which show that \mstar\ from \SL\ fits can be overestimated in the case of continuous SF by a factor of $\sim$2.5 at $t\sim$2 Gyr. This indicates that, whereas a broader spectral coverage improves estimates of \mstar\ for {\tt CONT} models by $\sim$20\%, the stellar mass is generally overestimated by \SL\ by no less than $\sim$50\% at these evolutionary stages.

	The counter-intuitive fact that the \2conf\ configuration, despite the omission of the nebular Balmer and Paschen jump, leads to a stronger discrepancy in the estimated \mstar\ (also in stellar age and metallicity; see below) suggests that incomplete flagging of weak emission lines that are not accounted for in the adopted spectral mask can strongly impact the spectral modelling result. Indeed, repetition of \SL\ fits on synthetic SEDs containing only the stellar and nebular continuum (obviously, a rather unrealistic set-up) has yielded an overall improvement in \mstar\ determinations for $\texttt{CONT}$ models and $t\leq 10^8$ yr by $\sim$0.25 dex in the case of \1conf, and up to $\sim$1 dex for \2conf\ (dotted lines labelled (c) in Fig.~\ref{Fig:COVER_-_Mstar_diff_vs_age}). It should be noted that fits in the latter set-up now yield a much better agreement than those in \1conf\ to the true \mstar, albeit a systematic overestimation by up to $\sim$1 dex for $t\leq 10^8$ yr and $\sim$0.15 dex for higher ages. This suggests that the residual nebular continuum
imposes an insurmountable limitation to purely stellar \PS\ tools even when a sophisticated procedure is adopted for a careful removal of emission lines, such as an initial coarse fit and subtraction of the stellar continuum for the sake of identification and subsequent masking of emission lines \citep[e.g.  ][]{Asari2007}.      
	
	As for \Fado, removal of hydrogen Balmer lines from the synthetic input SEDs forces the code to automatically cascade from its full-consistency mode (fitting both nebular lines and continuum) to its partial-consistency (fitting the nebular continuum only). This has virtually no effect for \1conf\ fits because the Balmer and Paschen jump holds in itself self-consistency constraints to the fit, whereas introducing slight uncertainties in \2conf\ models. Moreover, Figure \ref{Fig:BASEs_-_Mstar_diff_vs_age} shows that the \mstar\ trends found with both \Fado\ and \SL\ do not depend significantly on the adopted base library. With regard to other SF models, it can be seen from Fig. \ref{Fig:SFHs_-_Mstar_diff_vs_age} that stellar masses from \SL\ match those from \Fado\ within its typical uncertainty only for an age that is greater than 3 Gyr, depending on the assumed SFH.

% FIGURE  5 - - - - - - - - - - - - - - - - - - - - - - - - - - - - - - - - - - - - - - - - - - - - - - - - - - - - - - - - - - - - - - - - - - - - - - - - - - - - - - - - 
\begin{figure*}
\begin{center}
\includegraphics[width=0.99\textwidth]{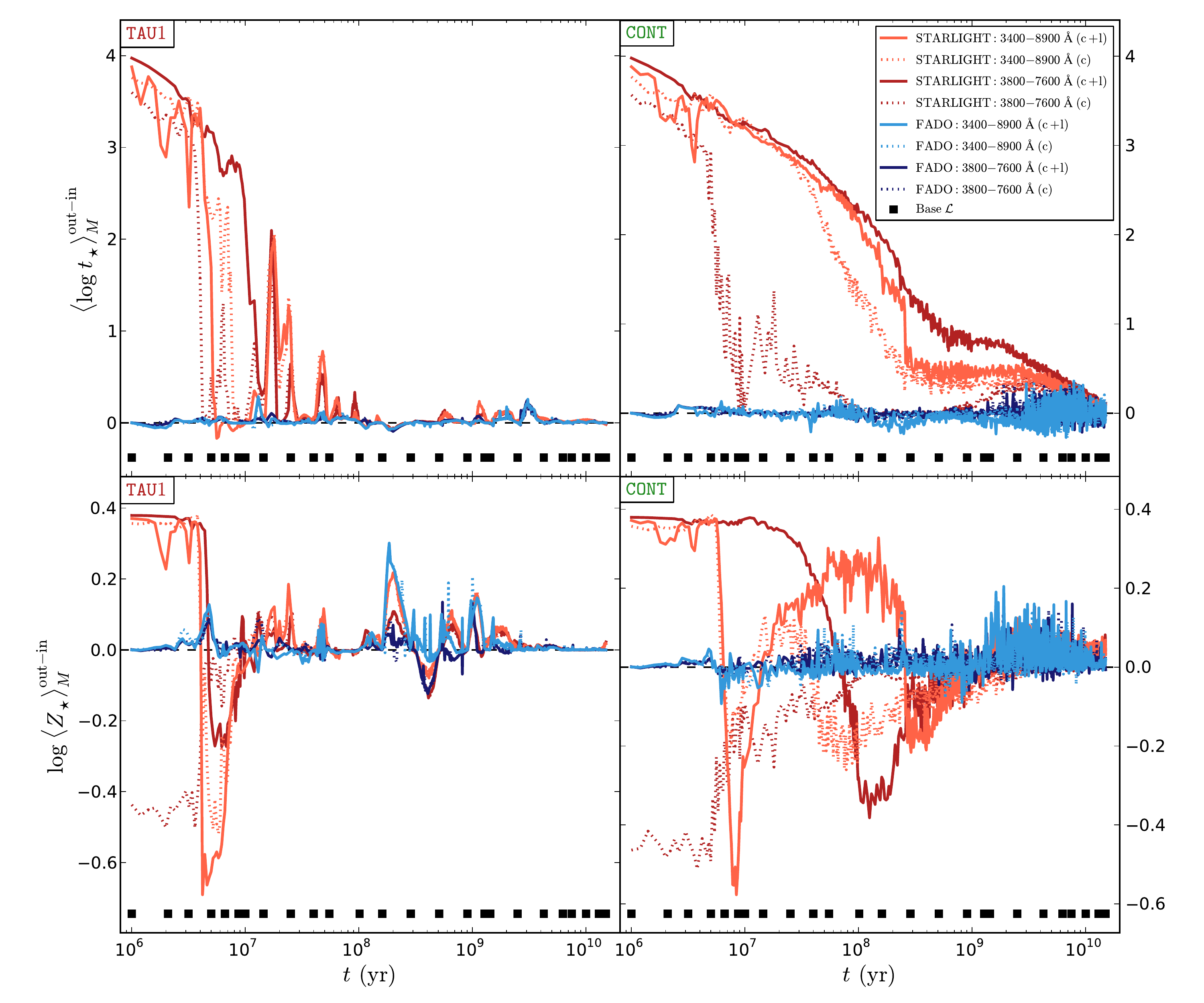}
\caption{Difference in the mass-weighted mean stellar age $\langle\log \, t_{\star}\rangle_M$ ({top row panels}) and metallicity $\log\langle Z_{\star}\rangle_M$ ({bottom row panels}) inferred from fits with \SL/\Fado\ ({out}) to synthetic SEDs from \Rebetiko\ ({in}) as a function of the CSP age $t$ for instantaneous burst ({left-hand side panels}) and continuous SFHS ({right-hand side panels}). Legend details are identical to those in Fig. \ref{Fig:COVER_-_Mstar_diff_vs_age}.}
\label{Fig:COVER_-_tM_diff_and_ZM_diff_vs_age}
\end{center}
\end{figure*}
% - - - - - - - - - - - - - - - - - - - - - - - - - - - - - - - - - - - - - - - - - - - - - - - - - - - - - - - - - - - - - - - - - - - - - - - - - - - - - - - - - - - - - - - - -

	Figure \ref{Fig:COVER_-_tM_diff_and_ZM_diff_vs_age} shows the difference in the mass-weighted mean stellar age \tm\ and metallicity \zm\ as a function of the CSP age $t$.  These results show that \Starlight\ overestimates \tm\ by up to 4 dex for young evolutionary stages, with the difference between output and input \tm\ showing both for $\texttt{TAU1}$ and $\texttt{CONT}$ a similar evolution with time as that seen in Fig. \ref{Fig:COVER_-_Mstar_diff_vs_age}.  For instance, for continuous SF and a model age of $\sim$3 Gyr, the \tm\ is overestimated with \SL\ fits by up to $\sim$0.5 dex. To the contrary, \Fado\ recovers within a precision of 0.1 dex, or better, the true \tm\ value over the entire interval  between 1 Myr and 15 Gyr, with only a few singular deviations of up to 0.3 dex at older evolutionary stages. As also apparent from Figs. \ref{Fig:BASEs_-_tM_diff_and_ZM_diff_vs_age} and \ref{Fig:SFHs_-_tM_diff_vs_age} for other base libraries and SFHs, respectively, \tm\ determinations with \SL\ and \Fado\ come to a match within 0.1 dex only at an age between 3 Gyr ({\tt DEL5}) and 8 Gyr ({\tt CONT}).

	Similarly, a comparison of the mass-weighted metallicity \zm\ attests to the capability of \Fado\ to recover the input value of $Z_{\odot}$ within 0.1-0.15 dex for both SFH scenarios and fitting set-ups, albeit a slight tendency to overestimate it by up to $\sim$0.2 dex for $t\ga 1$ Gyr.  The situation is more complex in the case of \SL. For instance, fits to $\texttt{TAU1}$ models indicate an over- or underestimation of \zm\ by up to $\sim$0.4 and 0.7 dex, respectively, for young ages ($t\leq 10$ Myr), with a gradually improving accuracy to $\la$0.15 dex for older ages.  As for continuous SF models, the \zm\ estimated with \SL\ shows a strong dependence on the fitted spectral range  and a reverse trend for the \1conf\ and \2conf\ set-up in different age intervals within the first 1 Gyr. Specifically, whereas fits within the extended spectral range \1conf\ underestimate \zm\ for 6--20 Myr and 0.3--1 Gyr, they yield an overestimate by up to $\sim$0.3 dex at intermediate ages between 20 Myr and 300 Myr. The opposite trend is seen for fits within the narrower interval \2conf, where \zm\ is underestimated by up $\sim$0.4 dex between 80 Myr and 400 Myr. The sensitivity of \zm\ on the considered spectral range is presumably due to the pronounced Balmer discontinuity that apparently forces \SL\ to select metal poor SSPs in an attempt to reproduce the steeply increasing SED continuum shortwards of 3650 \AA. The reason why \SL\ overestimates $Z_{\star}$ in \1conf\ for the ensuing $\sim$90 Myr and until the onset of the post-AGB phase is unclear, yet certainly unrelated to the construction of the SSP base, as apparent from Fig. \ref{Fig:BASEs_-_tM_diff_and_ZM_diff_vs_age}.  With regard to other SF models, Fig. \ref{Fig:SFHs_-_ZM_diff_vs_age} suggests that  \zm\ can be retrieved with \SL\  for certain SFHs (e.g. {\tt DEL4}, {\tt DEL5}) with a reasonably good accuracy ($\sim$0.15--0.2 dex) only for stellar populations older  than $\sim$1 Gyr and with $\mathrm{EW(H\alpha}) \la 60$ \AA.

% FIGURE  6 - - - - - - - - - - - - - - - - - - - - - - - - - - - - - - - - - - - - - - - - - - - - - - - - - - - - - - - - - - - - - - - - - - - - - - - - - - - - - - - - 
\begin{figure*}
\begin{center}
\includegraphics[width=0.99\textwidth]{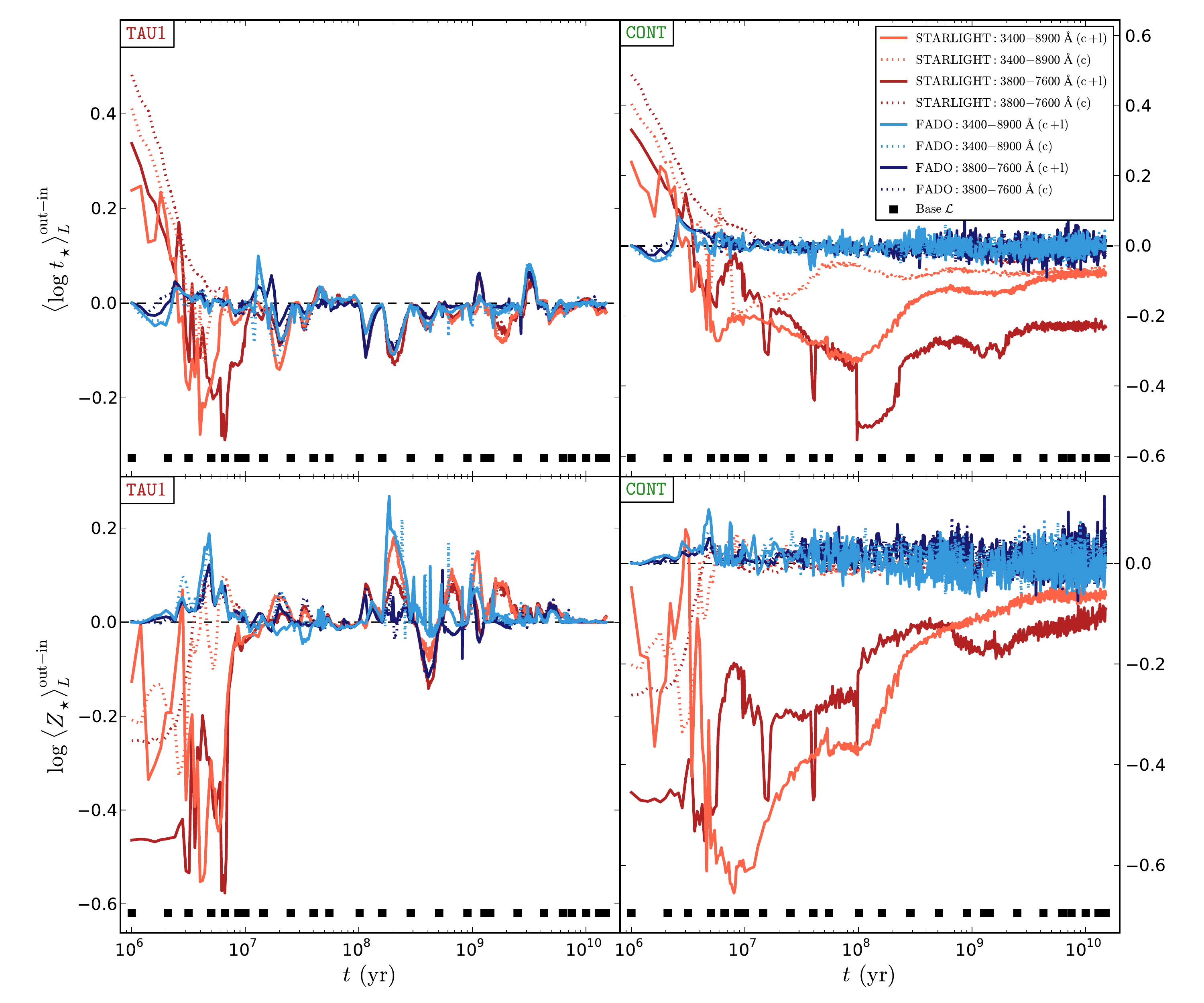}
\caption{Difference in the light-weighted mean stellar age $\langle\log \, t_{\star}\rangle_L$ ({top row panels}) and metallicity $\log\langle Z_{\star}\rangle_L$ ({bottom row panels}) inferred from fits with \SL/\Fado\ ({out}) to synthetic SEDs from \Rebetiko\ ({in}) as a function of the CSP age $t$ for instantaneous burst ({left-hand side panels}) and continuous SFHs ({right-hand side panels}). Legend details are identical to those in Fig. \ref{Fig:COVER_-_Mstar_diff_vs_age}.}
\label{Fig:COVER_-_tL_diff_and_ZL_diff_vs_age}
\end{center}
\end{figure*}
% - - - - - - - - - - - - - - - - - - - - - - - - - - - - - - - - - - - - - - - - - - - - - - - - - - - - - - - - - - - - - - - - - - - - - - - - - - - - - - - - - - - - - - - - -

	Figure \ref{Fig:COVER_-_tL_diff_and_ZL_diff_vs_age} shows the deviation of the light-weighted mean stellar age \tl\ and metallicity \zl\ from the  value of \Rebetiko\ input SEDs. In the case of $\texttt{TAU1}$, \Starlight\ tends to underestimate the stellar age by up to $\sim$0.3 dex between $\sim$3 and 10 Myr and by $\sim$0.55 dex for $\texttt{CONT}$ SEDs at $t \geq 10$ Myr.  A similar trend is apparent for \zl, which can be underestimated by between 0.6 and 0.05 dex, depending on age and fitted spectral range.  To the contrary, \Fado\ recovers in all cases both \tl\ and \zl\ within generally $\la$0.1 dex. The estimated metallicity for the instantaneous burst SFH displays an overestimation peak of up to $\sim$0.2 dex only at $\sim$200 Myr, which is to be attributed to the poor age coverage of \baseL. This is presumably also the reason for a slight age underestimation bump of up to $\sim$0.15 dex for $1 \lesssim t \lesssim 15$ Gyr in the case of {\tt CONT} models.

	Summarising, solutions with \SL\ are subject to substantial and complex biases that appear to be correlated with \ewha\ (i.e. the sSFR) and worsen by the inclusion of the Balmer jump that drives them towards a much too high contribution by young low-metallicity SSPs. The incomplete removal of weak emission lines with standard spectral masks is a secondary yet important cause for these biases. The latter could affect automated studies of large spectroscopic samples of star-forming galaxies (e.g. SDSS at $z\geq 0.05$), which are commonly done without an interactive control of the level of nebular continuum contamination and prominence of the Balmer jump.

	By contrast, \Fado\ does an overall reasonably good job in estimating the stellar mass, mean age, and mean metallicity regardless of the adopted fitting set-ups, SSP library, and age of the galaxy. The accuracy and robustness of determinations with \Fado\ results from its physical concept, in particular the estimate of the relative contribution of nebular continuum emission and associated mass fraction of ionising SSPs prior to full spectral decomposition into a linear combination of stellar SSPs. Evolutionary stages for which \Fado\ displays noticeable deviations from the true value (e.g. at $10\lesssim t \lesssim 30$ Myr) seem to be correlated with gaps in the age coverage of the SSP library used, as apparent from comparison of solutions obtained with \baseL\ (100 SSPs) with those from \baseF\ and \baseA\ (600 and 1326 SSPs, respectively) in Appendix \ref{Appendix:Bases}.

% !!!!!!!!!!!!!!!!!!!!!!!!!!!!!!!!!!!!!!!!!!!!!!!!!!!!!!!!!!!!!!!!!!!!!!!!!!!!!!!!!!!!!!!!!!!!!!!!!!!!!!!!!!!!!!!!!!!!!!!!!!!!!!!!!!!!!!!!!!!!!!!!!!!!!!!!!!!!!!!!!!!!!!!!!!!!!!!!!!!!!!!!!!!!!!!
\subsection{Hints at a bimodal SFH bias in purely stellar PS models}

% FIGURE  7 - - - - - - - - - - - - - - - - - - - - - - - - - - - - - - - - - - - - - - - - - - - - - - - - - - - - - - - - - - - - - - - - - - - - - - - - - - - - - - - - 
\begin{figure*}
\begin{center}
\includegraphics[width=0.99\textwidth]{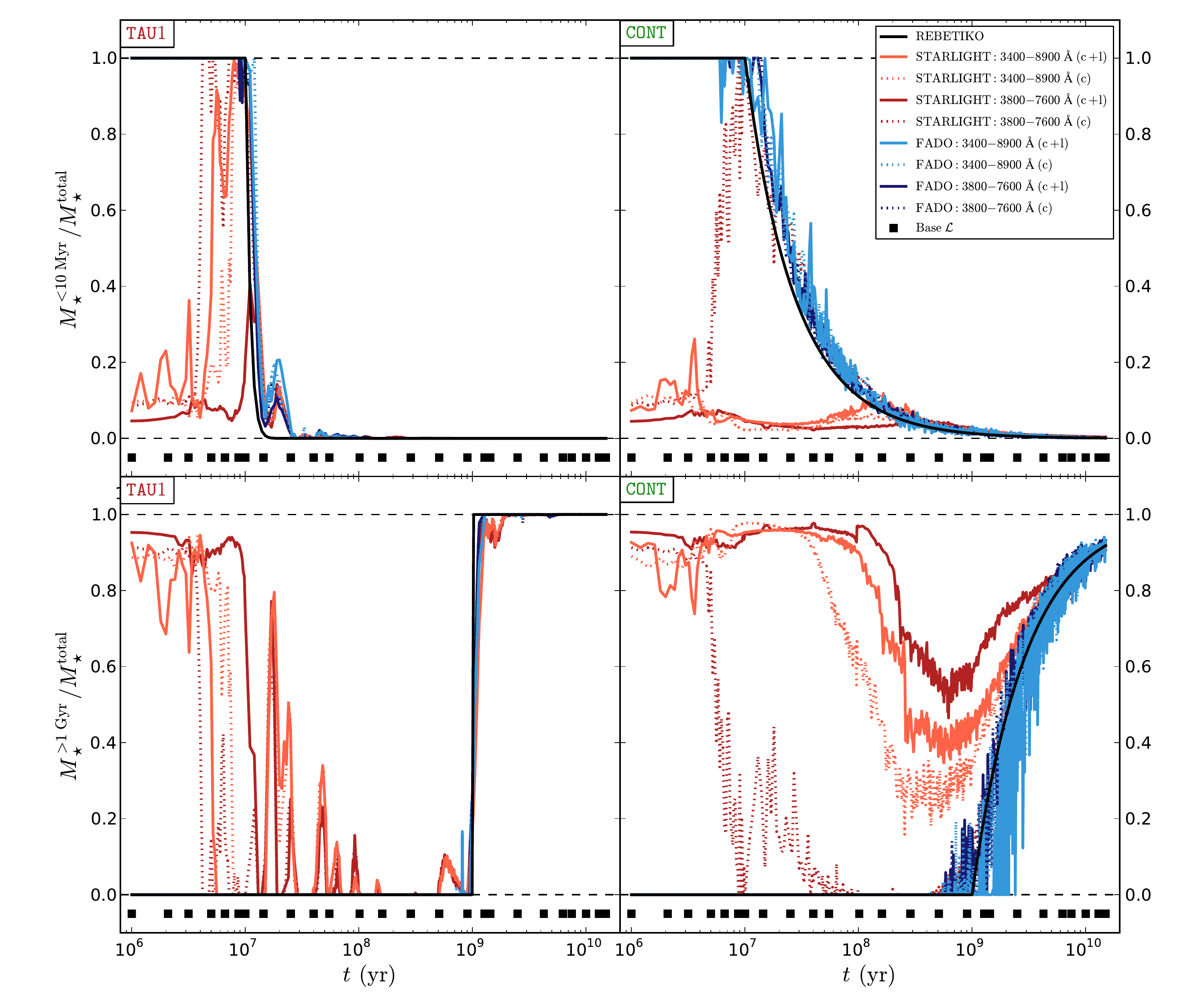}
\caption{Stellar mass fraction of SSPs younger than 10 Myr (\myoung; top row panels) and older than 1 Gyr (\mold; bottom row panels) as a function of the CSP age $t$ for instantaneous burst ({left-hand side panels}) and continuous SFHs ({right-hand side panels}). Black solid lines represent the true value of \myoung\ and \mold\ from \Rebetiko. Other legend details are identical to those in Fig. \ref{Fig:COVER_-_Mstar_diff_vs_age}.}
\label{Fig:COVER_-_young_and_old_parameters_vs_base_size}
\end{center}
\end{figure*}
% - - - - - - - - - - - - - - - - - - - - - - - - - - - - - - - - - - - - - - - - - - - - - - - - - - - - - - - - - - - - - - - - - - - - - - - - - - - - - - - - - - - - - - - - -

	In the foregoing section, we examined the output from the purely stellar PS code \SL\ with respect to mass- and light-weighted mean age and metallicity, following a usual approach in the validation of PS codes \citep[e.g.][]{CidFernandes2005}. However, it should be noted that such average quantities offer rather poor indicators for the ability of PS to recover key features of the mass assembly history of a galaxy.  For example, two substantially different SFHs: one involving continuous SF at a constant SFR and the other with two or several distinct SF episodes may yield the same \tl\ and \tm. Likewise, different CEHs may be indistinguishable from one another with respect to \zm\ and \zl. For this reason, a closer examination of PVs computed by modelling synthetic SEDs corresponding to simple parametric SFHs  appears to be a useful supplementary approach. This is also necessary for understanding the cause of the complex and in some cases contradictory biases documented for \SL\ in the previous section, as for example the simultaneous overestimation of \tm\ and underestimation of \tl\ (top panels of Figs. \ref{Fig:COVER_-_tM_diff_and_ZM_diff_vs_age} \& \ref{Fig:COVER_-_tL_diff_and_ZL_diff_vs_age}) during evolutionary phases with a high sSFR, thus strong nebular contamination (EW(H$\alpha$) $\ga200$ \AA; cf. Fig. \ref{Fig:SFHs_-_Mstar_diff_vs_age}).  The first bias, related to the overestimation of \mstar\ and \tm, was predicted in \citet{Izotov2011} and \citet{PapaderosOstlin2012}, who argued that the reddish nebular continuum of starburst galaxies forces purely stellar SED fitting codes to invoke a much too high fraction of old, high M/L ratio stars. In turn, this leads to an overestimation of both stellar mass and age, in agreement with the empirical evidence of Fig. \ref{Fig:SFHs_-_Mstar_diff_vs_age}. Whereas this bias was expected, this study is to our best knowledge where it has been first quantified by applying  a purely stellar PS code to synthetic SEDs with known physical and evolutionary properties.   The second bias, however, witnesses a trend that was not explored previously in sufficient detail, which is reminiscent of the report by \citet{Ocvirk2010} of ``fake star formation bursts'' registered in PVs computed with STECKMAP when fitting genuinely old stellar population spectra, such as globular clusters, namely a slight contamination of PVs from \Starlight\ by young SSPs, even for fits of evolved ($>$1 Gyr) CSPs. Whereas the origin of this effect is not entirely clear, the analysis by \citet{Ocvirk2010} suggests that it could partly originate from incompleteness in stellar libraries. We note that, because of its low-M/L ratio, this young and typically metal-poor stellar component may be important to or even dominate the optical luminosity, thereby leading to an underestimation of \tl\ and \zl.

	The seemingly contradictory evidence from Figs.~ \ref{Fig:COVER_-_Mstar_diff_vs_age} and \ref{Fig:COVER_-_tL_diff_and_ZL_diff_vs_age} suggests therefore a tendency of \SL\ to favour a bimodal solution for the SFH of star-forming galaxies. This potential caveat, even though partly blurred by inherent uncertainties in PS (e.g. age-metallicity-extinction degeneracy) and eventually somehow mitigated by a complete removal of emission lines, appears specially relevant during phases of high sSFR and in the light of our results deserves a closer examination. Clearly, an in-depth study of it is only possible by comparing the PVs from {\tt TAU1} and {\tt CONT} SFHs as determined with \SL, which is a task beyond the scope of the present article. However, as a preliminary test of the bimodality hypothesis, we study next the \mstar\ fraction of young ($\leq 10$ Myr) and old ($\geq 1$ Gyr) stars (\myoung\ and \mold, respectively) inferred from \Fado\ and \SL\ and compare their temporal evolution with that implied by the adopted SFHs.

	Figure \ref{Fig:COVER_-_young_and_old_parameters_vs_base_size} shows that \Fado\ closely matches the expected time evolution of \myoung\ and \mold\ for both SFHs and fitting set-ups over the entire time span between 1 Myr and 15 Gyr. This is not the case for \SL, where strong systematic deviations between the input and retrieved mass ratios are documented throughout. Indeed, the best agreement for an instantaneous burst model is achieved in the \1conf\ set-up at $t\sim$8 Myr, whereas fits in \2conf\ fall short of recovering \myoung\ for $t\leq 10$ Myr. Within this time interval and in particular for the \2conf\ configuration, the mass fractions \myoung\ and \mold\ estimated with \SL\ are nearly inverted, with \mold\ reaching up to 0.9 and {\it vice versa}. This offers a plausible explanation for the severe overestimation of \mstar\ and \tm. Another salient feature (bottom left panel) are strong singular peaks in \mold\ throughout the age interval 0.01--1 Gyr, which might be partly due to the sparse age coverage of the  SSP library \baseL. Quite importantly, \SL\ faithfully reproduces \myoung\ and \mold\ only at a post-starburst age of $\ga$2 Gyr. Even more complex is the situation in the case of a continuous SFH scenario. These models show that for $t\leq 100$ Myr, similar to {\tt TAU1} models, the mass ratios \myoung\ and \mold\ trends are inverted, with the latter exceeding $\sim$0.9 and thus erroneously implying a dominant old stellar population where there is none. This bias, which persists over the following few hundred Myr, naturally explains why \mstar\ is overestimated for {\tt CONT} models by a factor between $\sim$2 and $\sim$30 within the first 0.1--0.9 Gyr of galactic evolution (cf. Fig. \ref{Fig:COVER_-_Mstar_diff_vs_age}). Even for an age of 1 Gyr, where \mold=0, \SL\ fits indicate that between 40\% and 60\% of \mstar\ comes from older stars, where the \mold\ smoothly approaches the true value (black solid curve) with increasing age and matches it within the 1$\sigma$ dispersion of predictions with \Fado\ only after $t\ga 4$ Gyr.

% FIGURE  8 - - - - - - - - - - - - - - - - - - - - - - - - - - - - - - - - - - - - - - - - - - - - - - - - - - - - - - - - - - - - - - - - - - - - - - - - - - - - - - - - 
\begin{figure*}
\begin{center}
\includegraphics[width=0.99\textwidth]{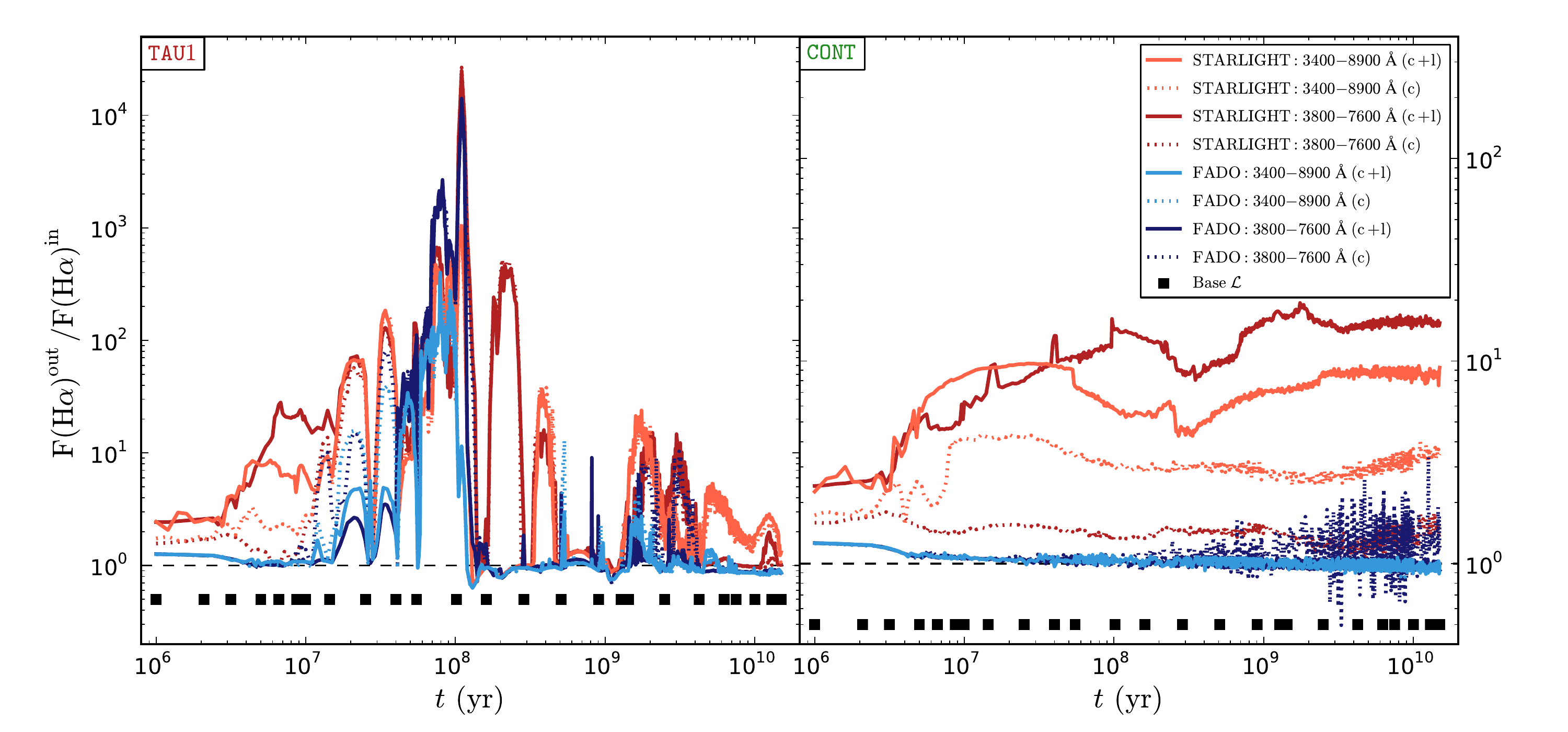}
\caption{Ratio of the H$\alpha$ flux \fha\ computed from the total \lyc\ photon output from the best-fitting PVs from \SL/\Fado\ (out) to the input value from \Rebetiko\ (in) as a function of the CSP age $t$ for instantaneous burst ({left-hand side panels}) and continuous SFHs ({right-hand side panels}).  Legend details are identical to those in Fig. \ref{Fig:COVER_-_Mstar_diff_vs_age}.}
\label{Fig:COVER_-_FHalpha_vs_base_size}
\end{center}
\end{figure*}
% - - - - - - - - - - - - - - - - - - - - - - - - - - - - - - - - - - - - - - - - - - - - - - - - - - - - - - - - - - - - - - - - - - - - - - - - - - - - - - - - - - - - - - - - -

	Summarising, the evidence of Fig. \ref{Fig:COVER_-_young_and_old_parameters_vs_base_size} supports the hypothesis that spectral fits with \SL\ tend towards a bimodal linear superposition of young and old SSPs during phases of significant sSFR (and, consequently, also nebular continuum contamination), which in extreme cases may be reduced into a two-component fit.  Because of the neglect of nebular continuum emission, hence also the lack of a mechanism ensuring consistency between the best-fitting  PV and \ne\ characteristics, the young-to-old \mstar\ ratio \citep[i.e. a proxy to the burst parameter; e.g.][]{Krueger1995}  is solely driven by $\chi^2$ minimisation in the SED continuum. The latter approach is known to result in degenerate solutions  \citep[e.g.][]{Guseva2001}. This concern is underpinned by the documented inversion of \myoung\ and \mold\ during phases  of high sSFR that erroneously indicates a dominant old stellar component where there is none. Consistent with the severely underestimated \myoung\ in phases of high sSFR is the slightly overestimated \myoung\ in the ensuing  phases of lower sSFR, which is visible on the top row panels of Fig. \ref{Fig:COVER_-_young_and_old_parameters_vs_base_size} all over $t\geq 10$ Myr ({\tt TAU1}) and between 0.01 and $\sim$2 Gyr ({\tt CONT}). The stagnation of \myoung\ to a non-zero level for {\tt TAU1} or its slight overestimation on timescales of several Gyr for {\tt CONT} further echoes the PV bimodality bias in \SL, thus explaining the underestimation of \tl.

	A possible excess in \myoung\ can also be tested through comparison of the H$\alpha$ flux \fha$^{\rm out}$ computed from PVs with the value  \fha$^{\rm in}$ corresponding to an assumed SFH. For this, we took a similar approach as in \citet{Papaderos2013} and \citet{Gomes2016a}, by first integrating the total Lyman continuum (\lyc) photon rate of the individual SSPs evaluated with a $\gamma_i >0$ and then converting it into \fha\ using standard photoionisation prescriptions while adopting standard gas conditions (i.e. $n_e = 100$ cm$^{-3}$ and $T_e = 10^4$ K ) and a zero \lyc\ escape fraction $f_{\rm esc}$. The H$\alpha$ luminosities were converted into fluxes assuming a uniform normalisation distance. Figure \ref{Fig:COVER_-_FHalpha_vs_base_size} shows as a function of time the ratio of \fha$^{\rm out}$ for PVs with \Fado\ and \SL\ to the theoretical value \fha$^{\rm in}$. In the case of {\tt TAU1} models, \Fado\ yields an overall good agreement with \fha$^{\rm in}$, except for the age interval 40--100 Myr where \fha$^{\rm out}$ locally exceeds \fha$^{\rm in}$ by up to a factor $\sim$300. The reason for this discrepancy is likely the poor coverage of the SSP base in the age interval prior to the onset of the post-AGB phase at $t=100$ Myr. Indeed, \baseL\ covers this 60 Myr interval, during which the \lyc\ rate declines by more than one dex \citep[cf. e.g.][]{CidFernandes2011,Gomes2016a} with only three SSPs  (50, 66, and 87 Myr; cf. Appendix \ref{Appendix:Bases}), which results in a very coarse estimate of the \lyc\ evolution. As for \SL, a discrepancy between \fha$^{\rm in}$ and \fha$^{\rm out}$ by a factor of $>$50 is present for $t\leq 40$ Myr, reaching a factor of $\geq$1000 at the post-AGB phase  around 100 Myr.  With regard to fits with \Fado\ of {\tt CONT} models, the right panel reveals an excellent agreement between \fha$^{\rm in}$ and \fha$^{\rm out}$ and has merely a minor deviation of $\la$20\% for $t\la 6$ Myr. Even though this is unsurprising, given that \Fado\ requires by design in its full-consistency fitting mode the best possible match between predicted and observed Balmer line luminosities, the quality and robustness of that match over $\geq$13 Gyr is reassuring.   To the contrary, the \fha$^{\rm out}$ corresponding to PVs with \SL\ exceeds the true value by a factor of a few throughout the considered age interval, which adds further support to the conclusion of a throughout slightly overestimated \myoung.

% !!!!!!!!!!!!!!!!!!!!!!!!!!!!!!!!!!!!!!!!!!!!!!!!!!!!!!!!!!!!!!!!!!!!!!!!!!!!!!!!!!!!!!!!!!!!!!!!!!!!!!!!!!!!!!!!!!!!!!!!!!!!!!!!!!!!!!!!!!!!!!!!!!!!!!!!!!!!!!!!!!!!!!!!!!!!!!!!!!!!!!!!!!!!!!!
% - - - - - - - - - - - - - - - - -- - - - - - - - - - - -  - - - -  DISCUSSION   - - - - - - - - - - - - - - - - - - - - - - - - - - - - - - - - - - - - - - - - 
% !!!!!!!!!!!!!!!!!!!!!!!!!!!!!!!!!!!!!!!!!!!!!!!!!!!!!!!!!!!!!!!!!!!!!!!!!!!!!!!!!!!!!!!!!!!!!!!!!!!!!!!!!!!!!!!!!!!!!!!!!!!!!!!!!!!!!!!!!!!!!!!!!!!!!!!!!!!!!!!!!!!!!!!!!!!!!!!!!!!!!!!!!!!!!!!
\section{Discussion}\label{Sec:Discussion}

	This is the first of a series of studies examining the capability of \PS\ codes to retrieve key physical and evolutionary properties of galaxies.  The tests presented for the two limiting cases of an instantaneous burst and continuous SF, along with  those provided in the Appendix \ref{Appendix:SFHs} for a broader range of parametric SFHs, attest to the capability of \Fado\ to estimate stellar quantities (e.g. total mass, mean age, and mean metallicity) with high accuracy (0.2 dex), even during evolutionary phases with a high sSFR and consequently severe nebular continuum contamination.

	It is worth reiterating that taking into account \ne\ is indispensable for a meaningful validation of any \PS\ code on realistic synthetic SEDs (cf. Sect. \ref{Sec:Introduction}). Although the approach taken in this work in this regard goes beyond those previously  adopted \citep[e.g.][]{CidFernandes2005,Chen2010,Wilkinson2017}, it has some important limitations to be kept in mind. For instance, the synthetic SEDs modelled both with \Fado\ and \SL\ correspond to the strongly simplified situation  of a star-forming galaxy whose entire \lyc\ output from stars is reprocessed into \ne\  by a gas reservoir of  constant electron temperature and density, where the stellar metallicity and intrinsic extinction are kept throughout fixed at $Z_{\odot}$ and 0~mag, respectively.  That such idealised conditions are highly improbable to be valid for the whole evolution of a galaxy seems reasonable. Furthermore, \Fado\ assumes in its standard full-consistency mode (FCmode; cf. GP17) that the \ne\  in a galaxy arises mainly from the stellar \lyc\ output. However, in the course of the evolution of the complex star-gas galaxy ecosystem, different constituents of the warm ISM may emerge and provide different contributions to the nebular SED across wavelength and galactocentric radius. These include, for instance, the diffuse ionised gas (DIG) or low-ionization nuclear emission-line region (LINER) emission that might be due to a diversity of gas excitation mechanisms, such as SF-driven winds, shocks, and photoionisation by post-AGB sources  \citep[e.g.][]{Binette1993,Allen2008,Stasinska2008,SharpBlandHawthorn2010,CidFernandes2011,Papaderos2013,Singh2013,Gomes2016a,Belfiore2017}. How these contributions might affect fits with \Fado\ in its  full-consistency mode FCmode needs to be investigated.

	Moreover, another unknown is the $f_{\rm esc}$ of \lyc\ radiation, which could strongly diminish the Balmer line fluxes and EWs that largely control the convergence of \Fado\ in its FCmode. For example, up to $\sim$50\% of the \lyc\ photons from massive stars could escape from HII regions \citep[e.g.][]{ER11} or low-mass SF galaxies \citep[e.g.][]{Izotov2018}. Even though the above \lyc\ leakers are to be considered extreme cases, and the current observational picture does not indicate a large $f_{\rm esc}$ for local star-forming galaxies,  the \lyc\ photon escape remains a potential source of uncertainty. As for the LINER nuclei of massive ETGs, the $f_{\rm esc}$ may exceed 0.9, thus placing \ne\ below the detection limit of standard spectral fitting codes \citep{Papaderos2013,Gomes2016a}\footnote{We note that the slightly overestimated fraction of young ionising stars in \SL\ fits (cf. Fig. \ref{Fig:COVER_-_FHalpha_vs_base_size}) does not alter our conclusions in these studies, since the high $f_{\rm esc}$ in ETG nuclei was inferred from analysis of the expected \lyc\ photon output from solely the evolved ($\geq$100 Myr) post-AGB stellar component.}. In the case of a virtually lineless galaxy, either because of a high $f_{\rm esc}$ or the absence of SF in a   passive galaxy, \Fado\ switches to the nebular continuum mode or purely stellar fitting mode (cf. GP17), dropping partially or completely the requirement of consistency between \ne\ and  SFH. 

	Another question concerns the impact of the featureless power-law continuum from an AGN on the SFH inferred from PS models.  \citet{Cardoso2016} presented a pilot study on this issue by simulating an AGN power-law component that is embedded within a \lyc-leaking ($f_{\rm esc}$=1) galaxy of 10 Gyr in age. This model set-up approximates the situation in many ETG LINER nuclei \citep{Papaderos2013,Gomes2016a} or high-\mstar\ bulges \citep{BredaPapaderos2018}. This study revealed that the AGN would generally evade detection with purely stellar \PS\ codes if its contribution to the monochromatic luminosity of the SED continuum at 4020 \AA\ were lower than $\sim$26\%. This and a subsequent study in \citet{Cardoso2017} have additionally shown that a buried AGN artificially rejuvenates a galaxy with respect to its \tl, whereas driving PVs towards a strongly overestimated \mstar\ and \tm, much like the biases documented in Sect. \ref{Sec:Results}. Clearly, an in-depth examination of this subject is worthwhile\footnote{We note that future public releases of \Fado\ will include the provision of an additional AGN power-law component.}. In which manner mechanisms other than photoionisation by stars (i.e. DIG, LINER, and AGN emission) could influence spectral fits with \Fado\ and other codes is largely unknown. For this reason, even though \Fado\ does a good job in fitting synthetic SEDs for the simplified model assumptions detailed in Sect. \ref{Sec:Methodology}, the results from this study should be viewed as an elementary test of \Fado\ and an illustration of biases that could affect automated studies of star-forming galaxies with \SL\ and similar purely stellar \PS\ codes in general. Nevertheless, the insights from Sect. \ref{Sec:Results} justify some preliminary remarks on the astrophysical implications of these biases and the gain in a self-consistent treatment of stellar and \ne\  in \PS.

{\bf i.} The fact that standard (purely stellar) \PS\ codes overestimate \mstar\ for high-sSFR galaxies, primarily as a consequence of the fact that the reddish nebular continuum drives fits towards a much too high contribution from old (high M/L) stars (cf. Sec \ref{Sec:Results}, e.g. \citealt{Izotov2011,PapaderosOstlin2012}), is especially important.   Depending on the method used to determine SFRs, this bias could lead to an underestimation of the sSFR and impact the slope of the \mstar\ versus SFR relation (i.e. the SFMS). A bias in \mstar\ also propagates into relations, such as \mstar\ versus sSFR \citep{Brinchmann2004} and \mstar\ versus its doubling time \citep{Noeske2007a}, which are commonly seen as manifestations of galaxy downsizing. Conversely, the galaxy downsizing itself, coupled with the \mstar\ overestimation in proportion to the sSFR, may further blur the picture of the galaxy assembly history and amplify observational biases in a  hardly predictable manner. As pointed out by \citet{PapaderosOstlin2012}, downsizing effects may further complicate the aforementioned biases, since \ne\ is expected to affect studies of galaxies of different mass over different timescales. Indeed, according to the downsizing picture of galaxy mass assembly,   lower-\mstar\ galaxies have a prolonged and significant SF activity at a late cosmic epoch when compared to more massive low-$z$ galaxies that assembled earlier and show virtually no \ne. An implication of this is an expected steepening of the SFMS relation (i.e. its $\alpha(z)$ exponent; cf. Sect. \ref{Sec:Introduction}) with decreasing redshift.

{\bf ii.} The over- or underestimation of the mass- and light-weighted mean stellar age, respectively, coupled with mass and metallicity biases, might also contribute to the scatter in scaling relations for star-forming galaxies such as age-metallicity or mass-metallicity relations. Furthermore, the overestimation of \tm\ for starburst galaxies, prompting their erroneous classification as old and massive, yet metal-poor systems,  could impact galaxy demographics in various respects. For instance, modelling of the optical rest-frame spectrum of a compact, high-$z$ starburst galaxy could result in its false classification as a massive, old, or young system (depending on whether the age is based on \tm\ or \tl) with an anomalously low metallicity. 

{\bf iii.} The tendency of \SL\ for a segregation of the best-fitting combination of SSPs into a bimodal (old {plus} young)  age distribution raises at least two questions. The first is whether this is a caveat of purely stellar \PS\ codes in general, eventually stemming from their neglect of \ne, thus also an inadequate treatment of emission-line infilling in stellar absorption features (e.g. H$\beta$ and H$\delta$ and other age-sensitive lines in the Lick system). It is certainly a task of considerable interest to understand this issue and its implications on the two-phase galaxy formation scenario it suggests.  The second question is whether or not the amplitude and duration of the two SF episodes in such a bimodal SFH are coupled. For instance, is the duration and/or the \mstar\ fraction produced in the initial and probably dominant SF phase inversely related to that in the second, more recent phase? Addressing this question is also important from the perspective of the co-evolution of super-massive black holes (SMBHs) and their galaxy hosts, since the former are believed to form early on and to regulate via AGN-driven feedback the ensuing stellar mass growth, as reflected in, for instance, a narrow range of values for the $M_{\rm SMBH}$/\mstar\ ratio \citep[$\approx 10^{-3}$;][]{KormendyHo2013}.  A possible inverse correlation of these two main SF phases would be in line with the conclusion that the SMBH controls the global SFH of galaxies in the sense of an enhanced mean SFR for systems that experienced a weaker initial SF phase and built an under-massive SMBH, and vice versa \citep[e.g.][]{MartinNavarro2018}. Further tests would be fundamental to address the contribution that \PS\ can make in solidifying this scenario.

{\bf iv.} Related to the SFH bimodality issue is presumably also the excess mass fraction by young ionising stars that is reflected in the over-predicted \ha\ luminosity from \SL\ fits. This substrate of young stellar populations, which is traceable in Fig. \ref{Fig:COVER_-_FHalpha_vs_base_size} for several Gyrs after the termination of a burst and throughout in the case of a continuous SF, is from the spectral modelling point of view an equivalent to ``SF frosting'' \citep{Trager2000,Kaviraj2009,Zibetti2017}. Whereas direct support for this scenario comes from UV and \ha\ studies of old and massive ETGs \citep[e.g.][]{SalimRich2010,Salim2012,Gomes2016c}, it should be kept in mind that a very low level 
of ongoing SF can be mimicked by \SL\ and possibly other purely stellar \PS\ codes \citep[cf.][]{Ocvirk2010}. 

	The brief remarks above arguably capture only minor aspects of the complexity and key astrophysical implications of spectral  synthesis and are merely meant to motivate further testsuits on increasingly realistic spectral models.

% !!!!!!!!!!!!!!!!!!!!!!!!!!!!!!!!!!!!!!!!!!!!!!!!!!!!!!!!!!!!!!!!!!!!!!!!!!!!!!!!!!!!!!!!!!!!!!!!!!!!!!!!!!!!!!!!!!!!!!!!!!!!!!!!!!!!!!!!!!!!!!!!!!!!!!!!!!!!!!!!!!!!!!!!!!!!!!!!!!!!!!!!!!!!!!!
% - - - - - - - - - - - - - - - - -- - - - - - - -  - -  - - CONCLUSIONS   - - - - - - - - - - - - - - - - - - - - - - - - - - - - - - - - - - - - - - -
% !!!!!!!!!!!!!!!!!!!!!!!!!!!!!!!!!!!!!!!!!!!!!!!!!!!!!!!!!!!!!!!!!!!!!!!!!!!!!!!!!!!!!!!!!!!!!!!!!!!!!!!!!!!!!!!!!!!!!!!!!!!!!!!!!!!!!!!!!!!!!!!!!!!!!!!!!!!!!!!!!!!!!!!!!!!!!!!!!!!!!!!!!!!!!!!
\section{Summary and conclusions}\label{Sec:Conclusions}

	This study constitutes a first step to a quantitative assessment of the capability of the spectral \PS\ code \Fado\ to recover key physical and evolutionary properties of galaxies.\ The \Fado\ code embodies a novel concept in spectral synthesis because it is the only publicly available \PS\ code that includes nebular continuum emission in spectral fits. Also this code ensures consistency between the \ne\ characteristics of a galaxy, such as hydrogen Balmer line luminosities and EWs as well as the shape of the SED around the Balmer and Paschen jump, with its best-fitting SFH and CEH. The tests presented in this work are based on modelling with \Fado\ and the purely stellar \PS\ code \SL\ of synthetic SEDs for nine parametric SFHs that are bracketed by the limiting cases of an instantaneous and continuous SF scenario.  A distinctive characteristic of this study as compared to most \PS\ validation testsuits before is that the synthetic SEDs modelled include \ne\ photoionised by the evolving stellar component and densely cover the age interval between 1 Myr and 15 Gyr. Moreover, this pilot study employs a number of simplifying assumptions for the modelled SEDs, such as standard conditions in the gas, constant stellar metallicity, and zero intrinsic extinction and \lyc\ photon escape fraction.  In total 716 SEDs for each SFH scenario were fitted with \Fado\ in two spectral intervals and using seven SSP libraries with 45 to 1326 SSPs. The same set of synthetic models was also modelled with four SSP bases with the code \SL, which may be regarded as representative of purely stellar PS tools. This task has allowed a first quantitative comparison of the results from self-consistent  and purely stellar \PS\ models to optical galaxy SEDs. We note that this study touches only briefly on the impact of fitted spectral range, construction of the SSP base, and quality   of emission-line flagging. These specifics in the fitting procedure, besides the spectral S/N ratio, are clearly important aspects to take into account for a conclusive comparison between different \PS\ codes in forthcoming articles. 

\smallskip
        The main results from this study may be summarised as follows:

{\bf i)} \Fado\ can recover from synthetic spectra the total mass and mean age and metallicity of stellar populations 
with a precision of typically better than 0.2 dex, even during evolutionary phases with strongly elevated sSFR and consequently severe contamination of SEDs by nebular line and continuum emission. This also applies to the relative mass fraction of young and old ($\leq$10 Myr and $\geq$1 Gyr, respectively) stellar populations,  which was found to closely match the time evolution implied by the assumed SFHs. The high fidelity with which \Fado\ recovers stellar population properties is also documented by the fact that the \ha\ luminosity implied by its best-fitting PVs shows an overall good agreement with the theoretically predicted values. Notwithstanding these facts, further work is needed for testing the code on more realistic synthetic SEDs that take into account intrinsic extinction, further ingredients of the warm ISM (e.g. DIG and shock-excited emission) and eventually an AGN continuum.

\smallskip
{\bf ii)} Modelling of synthetic SEDs with the purely stellar \PS\ code \SL\ has revealed a complex set of biases that appear to be primarily correlated with the level of nebular continuum contamination and also significantly depend on the specifics of the modelling procedure (e.g. flagging of weak emission prior to fitting, spectral range considered). In phases of strongly elevated sSFR the \mstar\ inferred from \SL\ models can surpass the true value by more than 1 dex. This is mainly because a purely stellar \PS\ code has no other option than to account for the reddish nebular continuum by invoking a much too high contribution from old high-M/L ratio SSPs. A consequence of this is the severe overestimation of the mass-weighted stellar age that can readily lead to the erroneous classification of a starburst galaxy as a massive old ETG. To the contrary, light-weighted stellar ages and metallicities from \SL\ can be substantially underestimated because the strong Balmer jump in high sSFR is compensated by invoking a low-metallicity young stellar component.   Furthermore, our analysis shows that the aforementioned biases in \SL\ generally persist, albeit becoming weaker, for at least the first 2--4 Gyrs of galactic evolution (depending on the assumed SFH) and until the \ewha\ has declined to levels below $\sim$60 \AA. Even though these biases depend on the spectral range fitted and become weaker when nebular line emission is completely excluded (i.e. when fitting the stellar and nebular continuum only), they pose a significant limitation to studies of star-forming galaxies with \SL\ and eventually also other purely stellar \PS\ codes.  Moreover, an intriguing feature of PVs from \SL\ for intermediate-to-high sSFR galaxies is their subtle tendency towards segregating into a bimodal stellar age distribution that appears to be consistent with a two-phase galaxy formation process. This artificial bimodality, which is documented through an inspection of the inferred-to-theoretical evolution of the \mstar\ fraction of stars younger than 10 Myr and older than 1 Gyr, is also reflected in an overestimated \lyc\ photon rate from young ($\leq$10 Myr) stars on timescales of $\sim$100 Myr and several Gyrs in the case of an instantaneous and continuous SF scenario, respectively. 

\smallskip
{\bf iii)} The biases in spectral models with \SL, if representative for purely stellar PS codes in general, could have important astrophysical implications. These include, among others, an increased dispersion or change of the slope of the SFMS and other scaling relations, such as the \mstar\ versus metallicity relation. Furthermore, a tendency for a SFH bimodality could be taken as supportive evidence for a correlated evolution between SMBHs and their galaxy hosts in the sense of an efficient suppression of SF at a late cosmic epoch due to an over-massive SMBH formed early on. Finally, the subtly overestimated mass fraction of a young ($\leq$10 Myr) stellar component on timescales of $\sim$100 Myr (several Gyrs) for an instantaneous (continuous) SF model could lend theoretical support to the SF frosting hypothesis. While all above hypotheses might be valid, a further rigorous investigation of \PS\ models using a suit of advanced tests on synthetic spectra appears fundamental for a better assessment of the contribution that spectral fitting could add towards their validation and refinement.

% !!!!!!!!!!!!!!!!!!!!!!!!!!!!!!!!!!!!!!!!!!!!!!!!!!!!!!!!!!!!!!!!!!!!!!!!!!!!!!!!!!!!!!!!!!!!!!!!!!!!!!!!!!!!!!!!!!!!!!!!!!!!!!!!!!!!!!!!!!!!!!!!!!!!!!!!!!!!!!!!!!!!!!!!!!!!!!!!!!!!!!!!!!!!!!!
% - - - - - - - - - - - - - - - - -- - - - - - - - - - - - - ACKNOWLEDGEMENTS   - - - - - - - - - - - - - - - - - - - - - - - - - - - - - - - 
% !!!!!!!!!!!!!!!!!!!!!!!!!!!!!!!!!!!!!!!!!!!!!!!!!!!!!!!!!!!!!!!!!!!!!!!!!!!!!!!!!!!!!!!!!!!!!!!!!!!!!!!!!!!!!!!!!!!!!!!!!!!!!!!!!!!!!!!!!!!!!!!!!!!!!!!!!!!!!!!!!!!!!!!!!!!!!!!!!!!!!!!!!!!!!!!

\begin{acknowledgements}

	We thank the anonymous referee for the valuable suggestions and comments. We also thank the European taxpayer, who in the spirit of solidarity and mutual respect between EU countries, provides the Funda\c{c}\~{a}o para a Ci\^{e}ncia e a Tecnologia (FCT) with a substantial fraction of the financial resources that allow it to sustain a research infrastructure in astrophysics in Portugal. Specifically, through European and national funding, we thank FCT that supported this work via Fundo Europeu de Desenvolvimento Regional (FEDER) through COMPETE2020 - Programa Operacional Competitividade e Internacionalização (POCI) by the grants UID/FIS/04434/2013 \& POCI-01-0145-FEDER-007672 and PTDC/FIS-AST/3214/2012 \& FCOMP-01-0124-FEDER-029170. 	
	%COMPETE by the grants UID/FIS/04434/2013 \& POCI-01-0145-FEDER-007672   and PTDC/FIS-AST/3214/2012 \& FCOMP-01-0124-FEDER-029170. 
	We acknowledge supported by European Community Programme (FP7/2007-2013) under grant agreement No. PIRSES-GA-2013-612701 (SELGIFS).	
	L.S.M.C. acknowledges funding by FCT through the grant CIAAUP-25/2018-BID in the context of the FCT project UID/FIS/04434/2013 \& POCI-01-0145-FEDER-007672.
	J.M.G. is supported by the fellowship CIAAUP-04/2016-BPD in the context of the FCT project UID/FIS/04434/2013 \& POCI-01-0145-FEDER-007672 and acknowledges the previous support by the fellowship SFRH/BPD/66958/2009 funded by FCT and POPH/FSE (EC).
	P.P. acknowledges support by FCT through Investigador FCT contract IF/01220/2013/CP1191/CT0002.
	We thank the members of the ``The assembly history of galaxies resolved in space and time''  Thematic Line of the Instituto de Astrof\' isica e Ci\^ encias do Espa\c co for a critical review of the manuscript and numerous valuable comments and suggestions.
	This research has made use of the NASA/IPAC Extragalactic Database (NED) which is operated by the Jet Propulsion Laboratory, California Institute of Technology, under contract with the National Aeronautics and Space Administration.    
\end{acknowledgements}

% !!!!!!!!!!!!!!!!!!!!!!!!!!!!!!!!!!!!!!!!!!!!!!!!!!!!!!!!!!!!!!!!!!!!!!!!!!!!!!!!!!!!!!!!!!!!!!!!!!!!!!!!!!!!!!!!!!!!!!!!!!!!!!!!!!!!!!!!!!!!!!!!!!!!!!!!!!!!!!!!!!!!!!!!!!!!!!!!!!!!!!!!!!!!!!!
% - - - - - - - - - - - - - - - - -- - - - - - - - - - - - - - - - - BIBLIOGRAPHY   - - - - - - - - - - - - - - - - - - - - - - - - - - - - - - -  - - -
% !!!!!!!!!!!!!!!!!!!!!!!!!!!!!!!!!!!!!!!!!!!!!!!!!!!!!!!!!!!!!!!!!!!!!!!!!!!!!!!!!!!!!!!!!!!!!!!!!!!!!!!!!!!!!!!!!!!!!!!!!!!!!!!!!!!!!!!!!!!!!!!!!!!!!!!!!!!!!!!!!!!!!!!!!!!!!!!!!!!!!!!!!!!!!!!
% BIBLIOGRAPHY ===============================================================
% Benchmark - Paper =========================================================

% BIBLIOGRAPHY ===============================================================

% !!!!!!!!!!!!!!!!!!!!!!!!!!!!!!!!!!!!!!!!!!!!!!!!!!!!!!!!!!!!!!!!!!!!!!!!!!!!!!!!!!!!!!!!!!!!!!!!!!!!!!!!!!!!!!!!!!!!!!!!!!!!!!!!!!!!!!!!!!!!!!!!!!!!!!!!!!!!!!!!!!!!!!!!!!!!!!!!!!!!!!!!!!!!!!!
% - - - - - - - - - - - - - - - - -- - - - - - - - - - - - - - - - -  - - - APPENDIX   - - - - - - - - - - - - - - - - - - - - - - - - - - - - - - -  - - - - 
% !!!!!!!!!!!!!!!!!!!!!!!!!!!!!!!!!!!!!!!!!!!!!!!!!!!!!!!!!!!!!!!!!!!!!!!!!!!!!!!!!!!!!!!!!!!!!!!!!!!!!!!!!!!!!!!!!!!!!!!!!!!!!!!!!!!!!!!!!!!!!!!!!!!!!!!!!!!!!!!!!!!!!!!!!!!!!!!!!!!!!!!!!!!!!!!
%!TEX encoding = UTF-8 Unicode\clearpage

% !!!!!!!!!!!!!!!!!!!!!!!!!!!!!!!!!!!!!!!!!!!!!!!!!!!!!!!!!!!!!!!!!!!!!!!!!!!!!!!!!!!!!!!!!!!!!!!!!!!!!!!!!!!!!!!!!!!!!!!!!!!!!!!!!!!!!!!!!!!!!!!!!!!!!!!!!!!!!!!!!!!!!!!!!!!!!!!!!!!!!!!!!!!!!!!
\appendix
\section[]{Comparison of FADO with STARLIGHT for different Bases}\label{Appendix:Bases}

	An important part of the testsuits presented in Sec. \ref{Sec:Methodology} focussed on the impact of the base library size on the estimation of stellar properties. Each base is composed of \cite{BruzualCharlot2003} SSPs with a \cite{Chabrier2003} IMF and Padova 1994 evolutionary tracks (\citealt{Alongi1993, Bressan1993, Fagotto1994a, Fagotto1994b, Girardi1996}). In total, seven bases with a dimension of between 45 and 1326 spectral templates were tested:
	
\begin{enumerate}
\item Base $\mathcal{N}$ with 45 SSPs, 15 ages ($t =$ 0.001, 0.00316, 0.00501, 0.01, 0.02512, 0.04, 0.10152, 0.28612, 0.64054, 0.90479, 1.434, 2.5, 5, 11 and 13 Gyr) and 3 metallicities ($Z =$ 0.004, 0.02 and 0.05); 

\item Base $\mathcal{L}$ with 100 SSPs,  25 ages ($t=$ 0.001, 0.00209, 0.00316, 0.00501, 0.00661, 0.00871, 0.01, 0.01445, 0.02512, 0.04, 0.055, 0.10152, 0.1609, 0.28612, 0.5088, 0.90479, 1.27805, 1.43400, 2.5, 4.25, 6.25, 7.5, 10, 13 and 15 Gyr) and 4 metallicities ($Z = 0.004$, 0.00800, 0.02 and 0.05); 

\item Base $\mathcal{S}$ with 150 SSPs, 25 ages ($t =$ 0.001, 0.00316, 0.00501, 0.00661, 0.00871, 0.01, 0.01445, 0.02512, 0.04, 0.055, 0.10152, 0.1609, 0.28612, 0.5088, 0.90479, 1.27805, 1.434, 2.5, 4.25, 6.25, 7.5, 10, 13, 15 and 18 Gyr) and 6 metallicities ($Z = 0.0001$, 0.0004, 0.004, 0.008, 0.02 and 0.05);

\item Base $\mathcal{J}$ with 162 SSPs, 27 ages ($t =$ 0.001, 0.00126, 0.002, 0.00316, 0.00501, 0.00661, 0.00871, 0.01, 0.01445, 0.02512, 0.04, 0.055, 0.10152, 0.1609, 0.28612, 0.5088, 0.90479, 1.27805, 1.434, 2.5, 4.25, 6.25, 7.5, 10, 13, 15 and 18 Gyr) and 6 metallicities ($Z = 0.0001$, 0.0004, 0.004, 0.008, 0.02 and 0.05);

\item Base $\mathcal{K}$ with 360 SSPs, 60 ages ($t =$ 0.001, 0.00105, 0.0011, 0.00115, 0.0012, 0.00126, 0.00132, 0.00138, 0.00145, 0.00151, 0.00158, 0.00166, 0.00174, 0.00182, 0.00191, 0.002, 0.00209, 0.00219, 0.00229, 0.0024, 0.00251, 0.00263, 0.00275, 0.00288, 0.00302, 0.00316, 0.00331, 0.00347, 0.00363, 0.0038, 0.00398, 0.00417, 0.00437, 0.00457, 0.00479, 0.00501, 0.00661, 0.00871, 0.01, 0.01445, 0.01905, 0.02512, 0.04, 0.055, 0.10152, 0.1609, 0.28612, 0.5088, 0.90479, 1.27805, 1.434, 2.5, 4.25, 6.25, 7.5, 10, 13, 13.75, 15 and 18 Gyr) and 6 metallicities ($Z = 0.0001$, 0.0004, 0.004, 0.008, 0.02 and 0.05);

\item Base $\mathcal{F}$ with 600 SSPs, 100 ages ($t =$ 0.001, 0.00105, 0.0011, 0.00115, 0.0012, 0.00126, 0.00132, 0.00138, 0.00145, 0.00151, 0.00158, 0.00166, 0.00174, 0.00182, 0.00191, 0.002, 0.00209, 0.00219, 0.00229, 0.0024, 0.00251, 0.00263, 0.00275, 0.00288, 0.00302, 0.00316, 0.00331, 0.00347, 0.00363, 0.0038, 0.00398, 0.00417, 0.00437, 0.00457, 0.00479, 0.00501, 0.00525, 0.0055, 0.00575, 0.00603, 0.00631, 0.00661, 0.00692, 0.00724, 0.00759, 0.00794, 0.00832, 0.00871, 0.00912, 0.00955, 0.01, 0.01047, 0.01096, 0.01148, 0.01202, 0.01259, 0.01318, 0.0138, 0.01445, 0.01514, 0.01585, 0.0166, 0.01738, 0.0182, 0.01905, 0.01995, 0.02089, 0.02188, 0.02291, 0.02399, 0.02512, 0.031, 0.04, 0.0475, 0.055, 0.07187, 0.08064, 0.10152, 0.1278, 0.1609, 0.20256, 0.255, 0.32103, 0.3602, 0.5088, 0.7187, 0.90479, 1.01519, 1.27805, 1.434, 1.8, 2.2, 3, 4.25, 5, 6, 7, 9, 11.25 and 13.5 Gyr) and 6 metallicities ($Z = 0.0001$, 0.0004, 0.004, 0.008, 0.02 and 0.05);

\item Base $\mathcal{A}$ with 1326 SSPs, 221 ages and 6 metallicites. This base contains all SSPs with \cite{Chabrier2003} IMF and Padova 1994 evolutionary tracks (\citealt{Alongi1993, Bressan1993, Fagotto1994a, Fagotto1994b, Girardi1996}). 

\end{enumerate}

	Both bases $\mathcal{N}$ and $\mathcal{S}$ are included in the distribution package of \Starlight. Moreover, $\mathcal{L}$ is similar to that adopted by \cite{Asari2007}, although with a slightly better coverage of young evolutionary stages and without SSPs with a metallicity $Z = 0.0001$ and 0.0004. Whereas tests with \Fado\ were carried out with all seven aforementioned bases, spectral modelling with \SL\ was restricted to the base libraries $\mathcal{N}$, $\mathcal{L}$, $\mathcal{S,}$ and $\mathcal{J}$, given the restriction of this code to a maximum number of 300 spectral elements in its base library.

\begin{figure*}
\begin{center}
\includegraphics[width=0.99\textwidth]{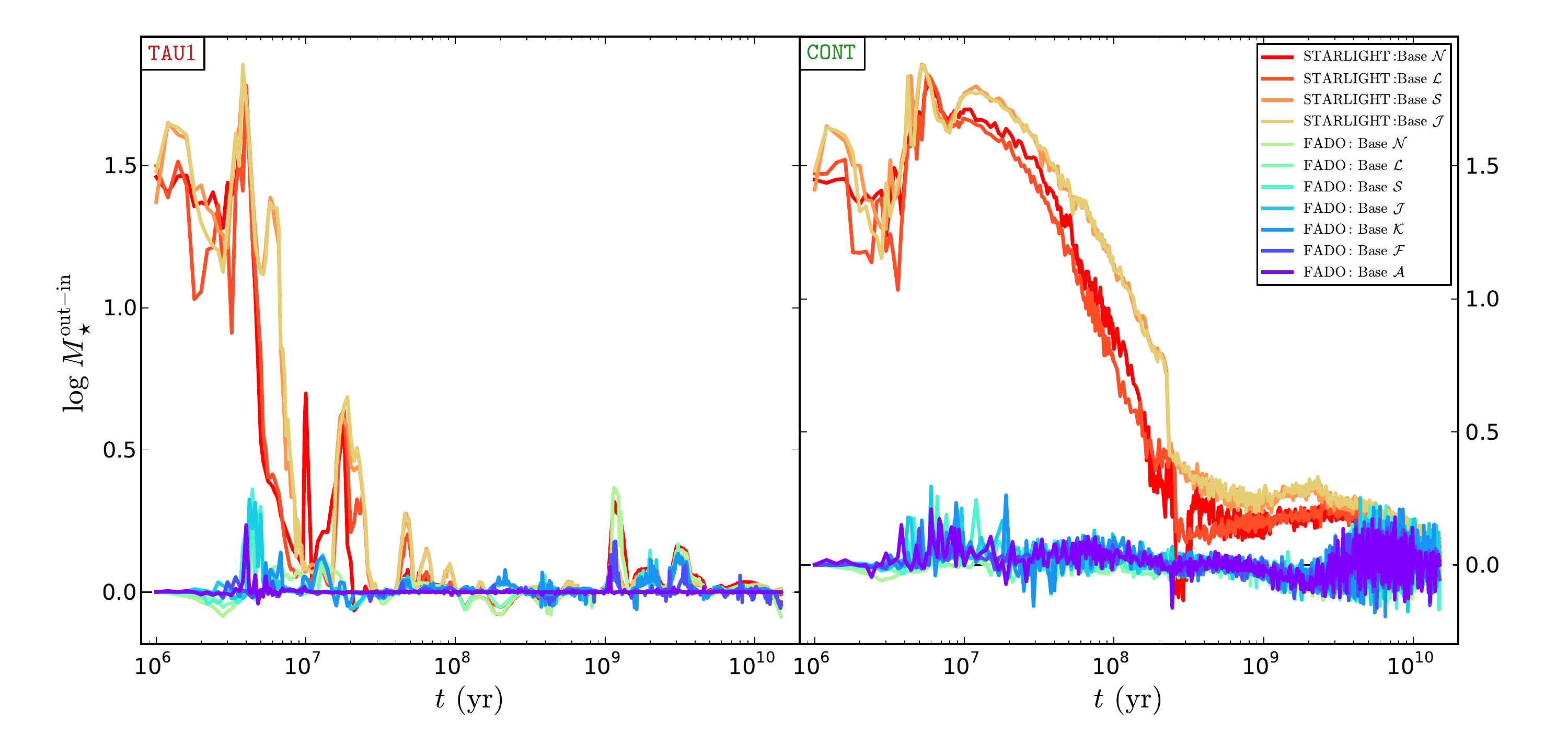}
\caption{Difference in the total stellar mass $M_{\star}$ between \Starlight/\Fado\ ({out}) and \Rebetiko\ ({in}) values as a function of the CSP age $t$ for instantaneous burst ({left panel}) and continuous SFHs ({right panel}). Check legend for more details.}
\label{Fig:BASEs_-_Mstar_diff_vs_age}
\end{center}
\end{figure*}

\begin{figure*}
\begin{center}
\includegraphics[width=0.99\textwidth]{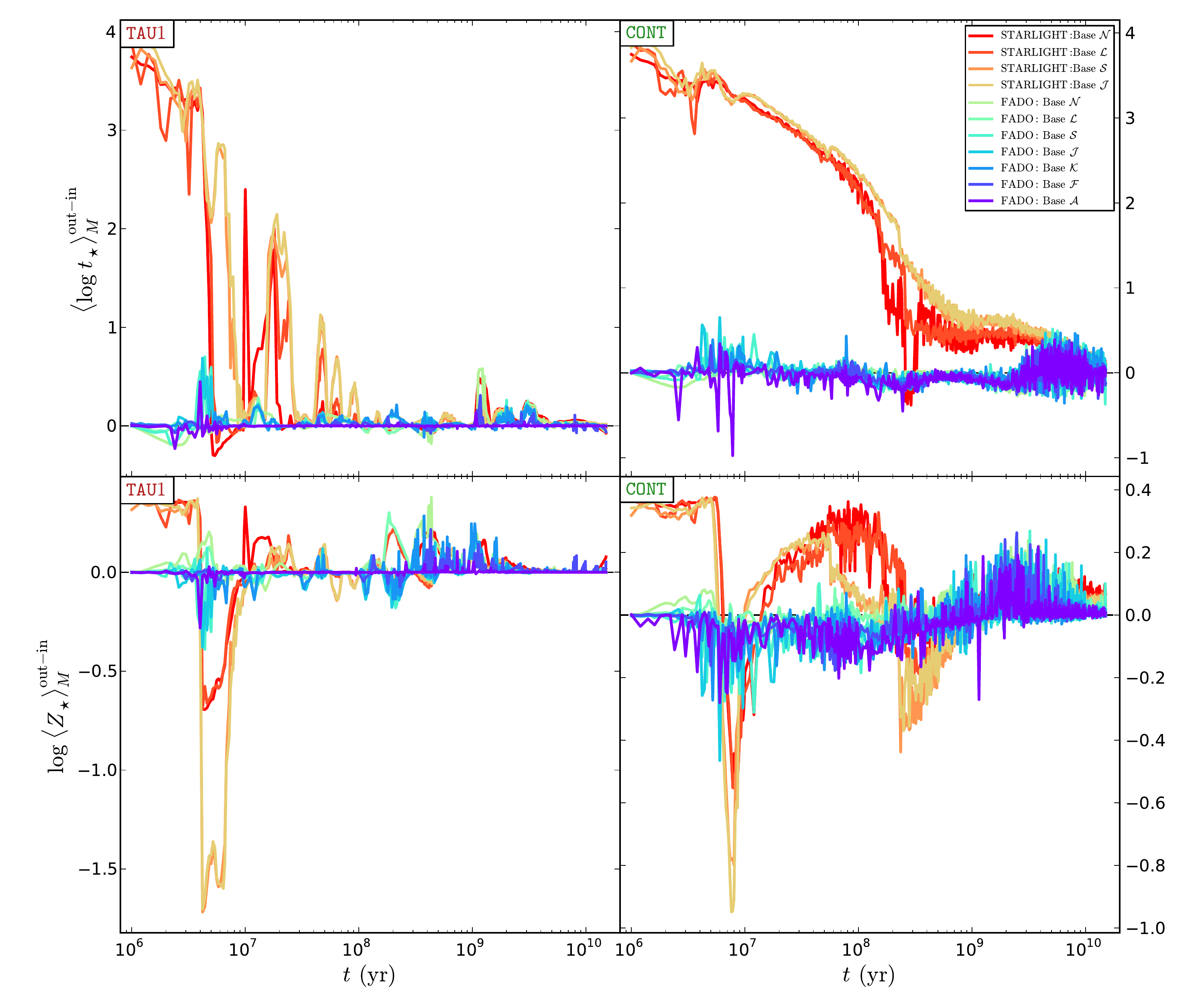}
\caption{Difference in the mass-weighted mean stellar age $\langle\log \, t_{\star}\rangle_M$ ({top row panels}) and  mean stellar metallicity $\log\langle Z_{\star}\rangle_M$ ({bottom row panels}) between \Starlight/\Fado\ ({out}) and \Rebetiko\ ({in}) values as a function of the CSP age $t$ for instantaneous burst ({left-hand side panels}) and continuous SFHs ({right-hand side panels}). Check legend for more details. }
\label{Fig:BASEs_-_tM_diff_and_ZM_diff_vs_age}
\end{center}
\end{figure*}

\begin{figure*}
\begin{center}
\includegraphics[width=0.99\textwidth]{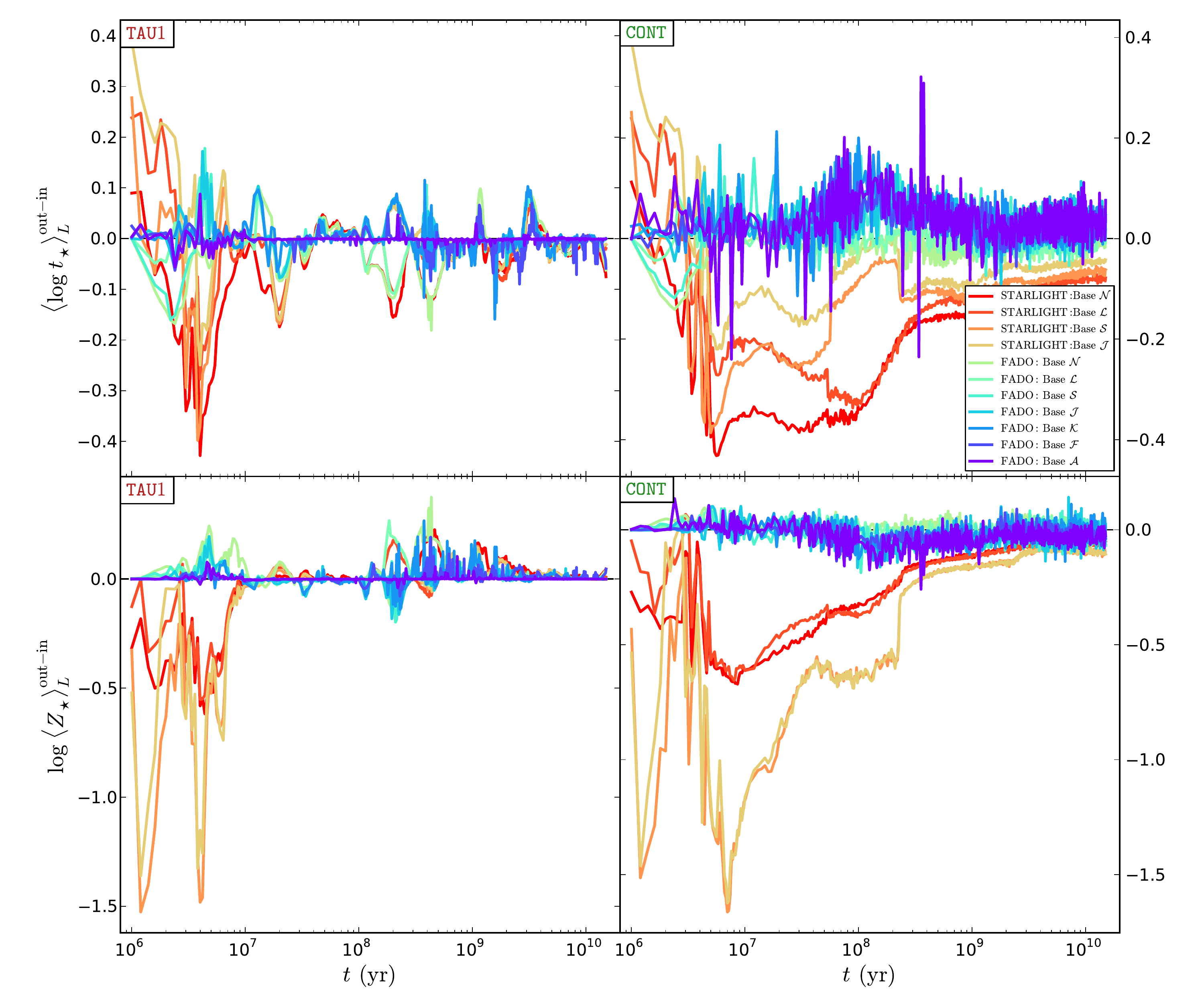}
\caption{Difference in the light-weighted mean stellar age $\langle\log \, t_{\star}\rangle_L$ ({top row panels}) and  mean stellar metallicity $\log\langle Z_{\star}\rangle_L$ ({bottom row panels}) between \Starlight/\Fado\ ({out}) and \Rebetiko\ ({in}) values as a function of the CSP age $t$ for instantaneous burst ({left-hand side panels}) and continuous SFHs ({right-hand side panels}). Check legend for more details. }
\label{Fig:BASEs_-_tL_diff_and_ZL_diff_vs_age}
\end{center}
\end{figure*}

	Overall, results for the total stellar mass and mean stellar age and metallicity using fitting set-up \1conf\ and presented in Figs. \ref{Fig:BASEs_-_Mstar_diff_vs_age}--\ref{Fig:BASEs_-_tL_diff_and_ZL_diff_vs_age} show uncertainties and trends similar to those shown in Figs. \ref{Fig:COVER_-_Mstar_diff_vs_age}--\ref{Fig:COVER_-_tL_diff_and_ZL_diff_vs_age}. In fact, Figs. \ref{Fig:BASEs_-_Mstar_diff_vs_age}--\ref{Fig:BASEs_-_tL_diff_and_ZL_diff_vs_age} suggest that an increasing number of SSPs in the base library does not improve significantly the stellar properties estimates. The most important feature seems to be the adequacy of the age and metallicity coverage of a base to a specific input spectrum or set of spectra. Whereas a closer examination of this issue is certainly needed, the results presented in this study suggest that \baseL\ yields a good compromise between the accuracy with which the most important stellar properties are estimated and the computational time required for the spectral modelling.

% FIGURE  A1 - - - - - - - - - - - - - - - - - - - - - - - - - - - - - - - - - - - - - - - - - - - - - - - - - - - - - - - - - - - - - - - - - - - - - - - - - - - - - - - 
% - - - - - - - - - - - - - - - - - - - - - - - - - - - - - - - - - - - - - - - - - - - - - - - - - - - - - - - - - - - - - - - - - - - - - - - - - - - - - - - - - - - - - - - - -
% FIGURE  A2 - - - - - - - - - - - - - - - - - - - - - - - - - - - - - - - - - - - - - - - - - - - - - - - - - - - - - - - - - - - - - - - - - - - - - - - - - - - - - - - 
% - - - - - - - - - - - - - - - - - - - - - - - - - - - - - - - - - - - - - - - - - - - - - - - - - - - - - - - - - - - - - - - - - - - - - - - - - - - - - - - - - - - - - - - - -
% FIGURE  A3 - - - - - - - - - - - - - - - - - - - - - - - - - - - - - - - - - - - - - - - - - - - - - - - - - - - - - - - - - - - - - - - - - - - - - - - - - - - - - - - 
% - - - - - - - - - - - - - - - - - - - - - - - - - - - - - - - - - - - - - - - - - - - - - - - - - - - - - - - - - - - - - - - - - - - - - - - - - - - - - - - - - - - - - - - - -

%\clearpage
% !!!!!!!!!!!!!!!!!!!!!!!!!!!!!!!!!!!!!!!!!!!!!!!!!!!!!!!!!!!!!!!!!!!!!!!!!!!!!!!!!!!!!!!!!!!!!!!!!!!!!!!!!!!!!!!!!!!!!!!!!!!!!!!!!!!!!!!!!!!!!!!!!!!!!!!!!!!!!!!!!!!!!!!!!!!!!!!!!!!!!!!!!!!!!!!
\section[]{Comparison of FADO with STARLIGHT for different SFHs}\label{Appendix:SFHs}

\begin{figure*}
\begin{center}
\includegraphics[width=0.99\textwidth]{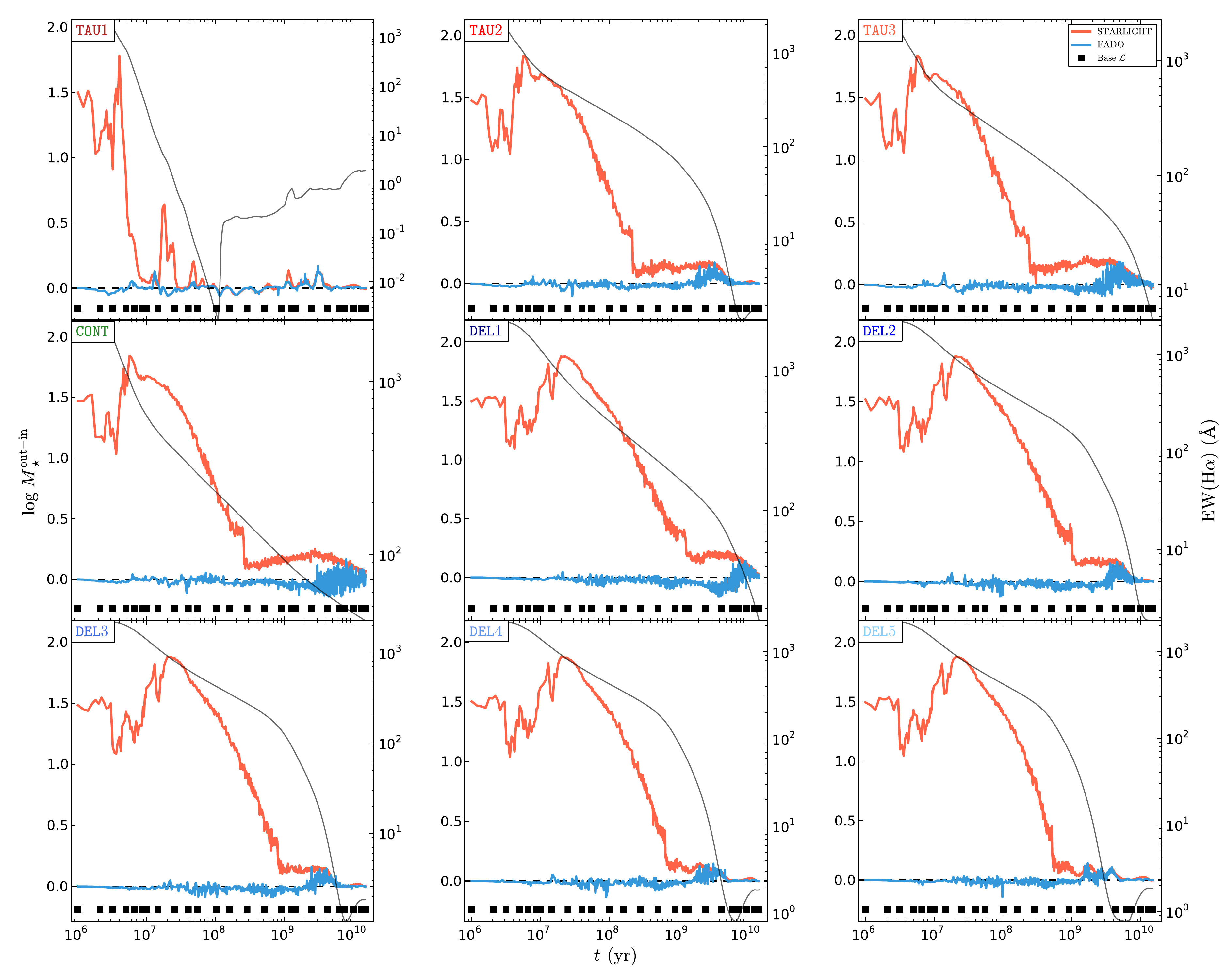}
\caption{Difference in the total stellar mass $M_{\star}$ between \Starlight/\Fado\ ({out}) and \Rebetiko\ ({in}) values as a function of the CSP age $t$ for different SFHs (see \citealt{Cardoso2017} for more details), from {left-} to {right-hand side} and {top} to {bottom}: 3 exponentially declining (\texttt{TAU1}, \texttt{TAU2} and \texttt{TAU3}), 1 continuous (\texttt{CONT}) and 5 delayed (\texttt{DEL1}, \texttt{DEL2}, \texttt{DEL3}, \texttt{DEL4} and \texttt{DEL5}). Red and blue lines represent results with \Starlight\ and \Fado, respectively. The black solid line represents the H$\alpha$ EW value (see right-hand side ordinate) and black solid squares show the age coverage of the adopted base library \baseL. }
\label{Fig:SFHs_-_Mstar_diff_vs_age}
\end{center}
\end{figure*}

\begin{figure*}
\begin{center}
\includegraphics[width=0.99\textwidth]{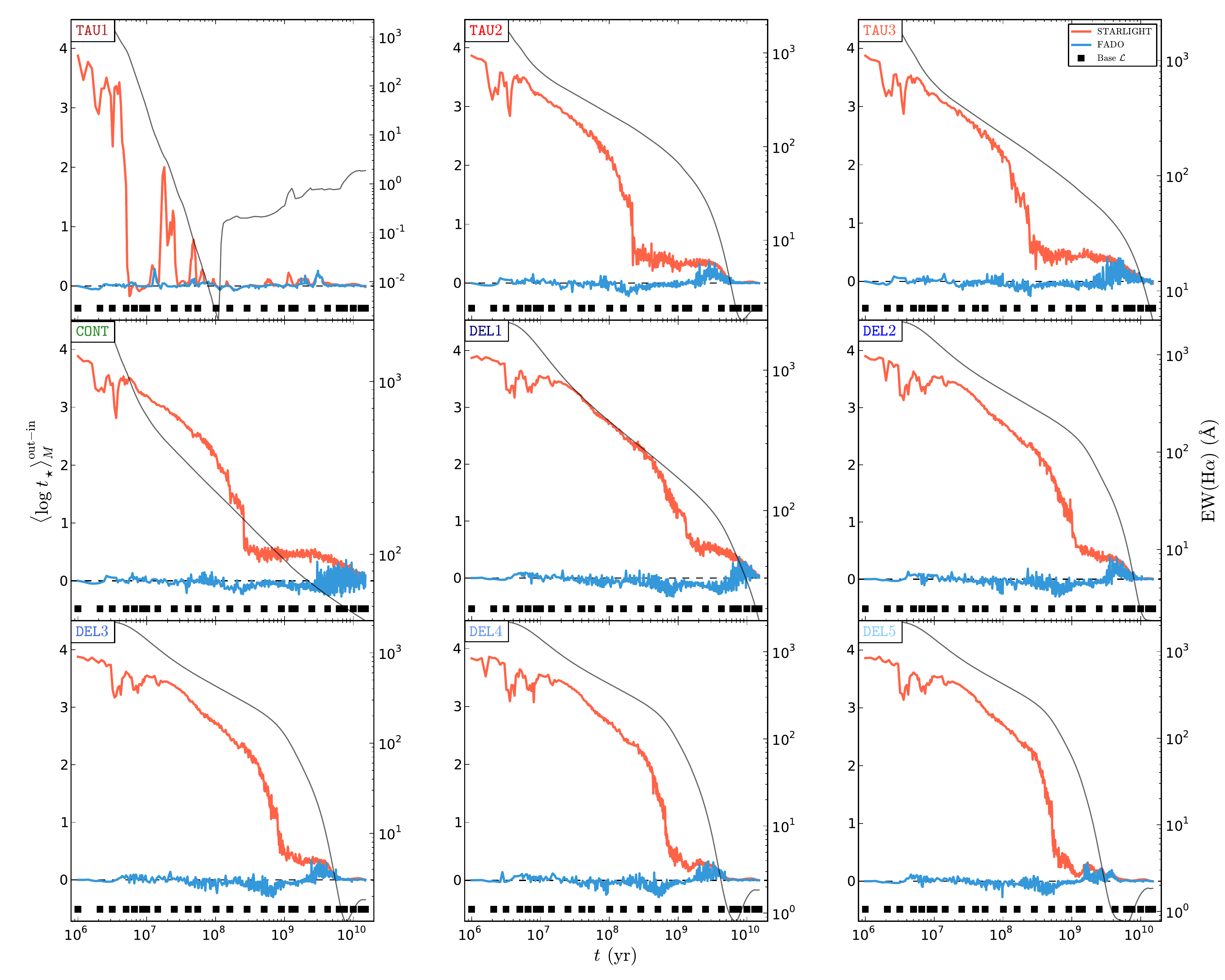}
\caption{Difference in the mass-weighted mean stellar age $\langle\log \, t_{\star}\rangle_M$ between \Starlight/\Fado\ ({out}) and \Rebetiko\ ({in}) values as a function of the CSP age $t$ for different SFHs. Legend details are identical to those in in Figure \ref{Fig:SFHs_-_Mstar_diff_vs_age}. }
\label{Fig:SFHs_-_tM_diff_vs_age}
\end{center}
\end{figure*}

\begin{figure*}
\begin{center}
\includegraphics[width=0.99\textwidth]{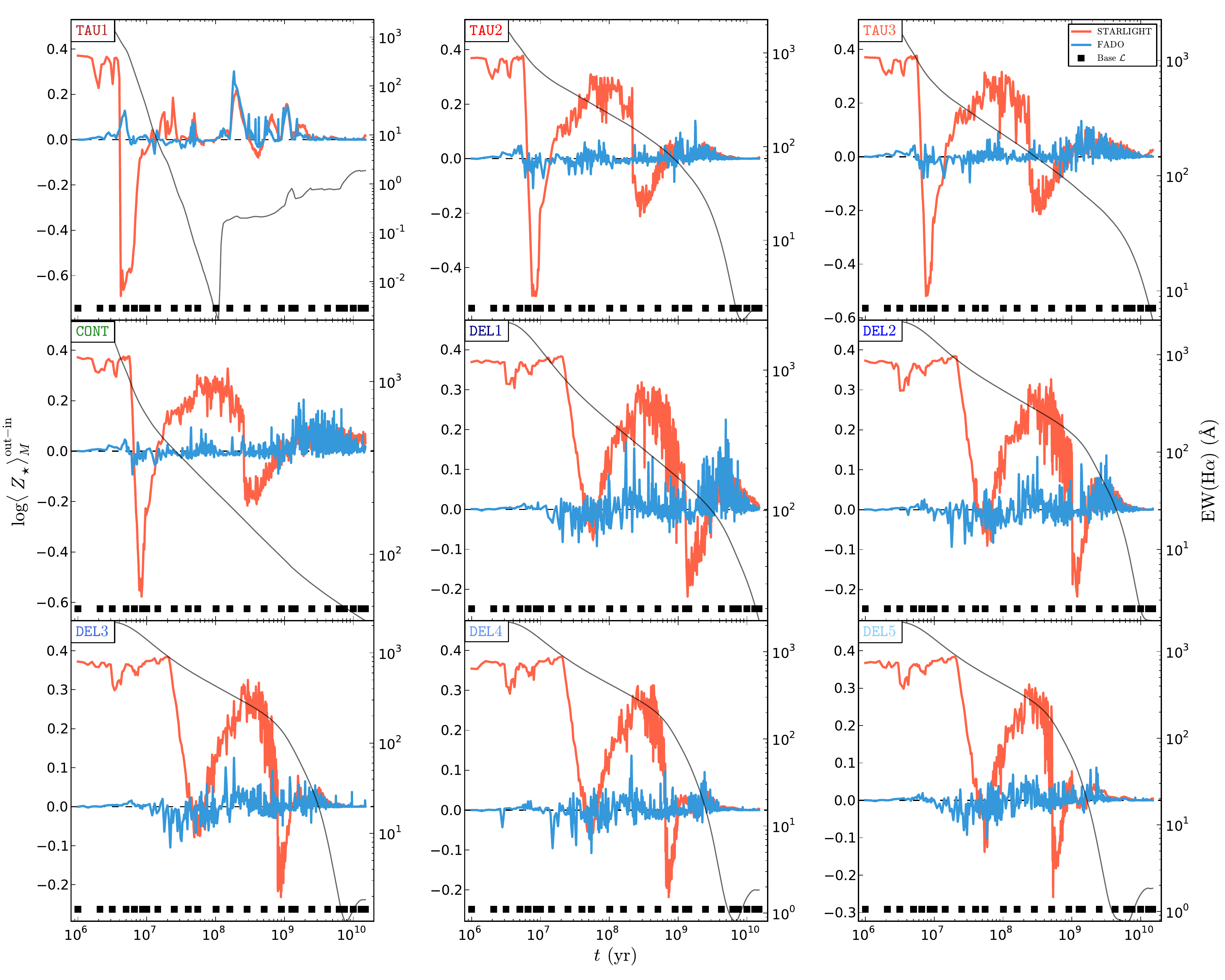}
\caption{Difference in the mass-weighted mean stellar metallicity $\log \langle Z_{\star}\rangle_M$ between \Starlight/\Fado\ ({out}) and \Rebetiko\ ({in}) values as a function of the CSP age $t$ for different SFHs. Legend details are identical to those in in Figure \ref{Fig:SFHs_-_Mstar_diff_vs_age}. }
\label{Fig:SFHs_-_ZM_diff_vs_age}
\end{center}
\end{figure*}

\begin{figure*}
\begin{center}
\includegraphics[width=0.99\textwidth]{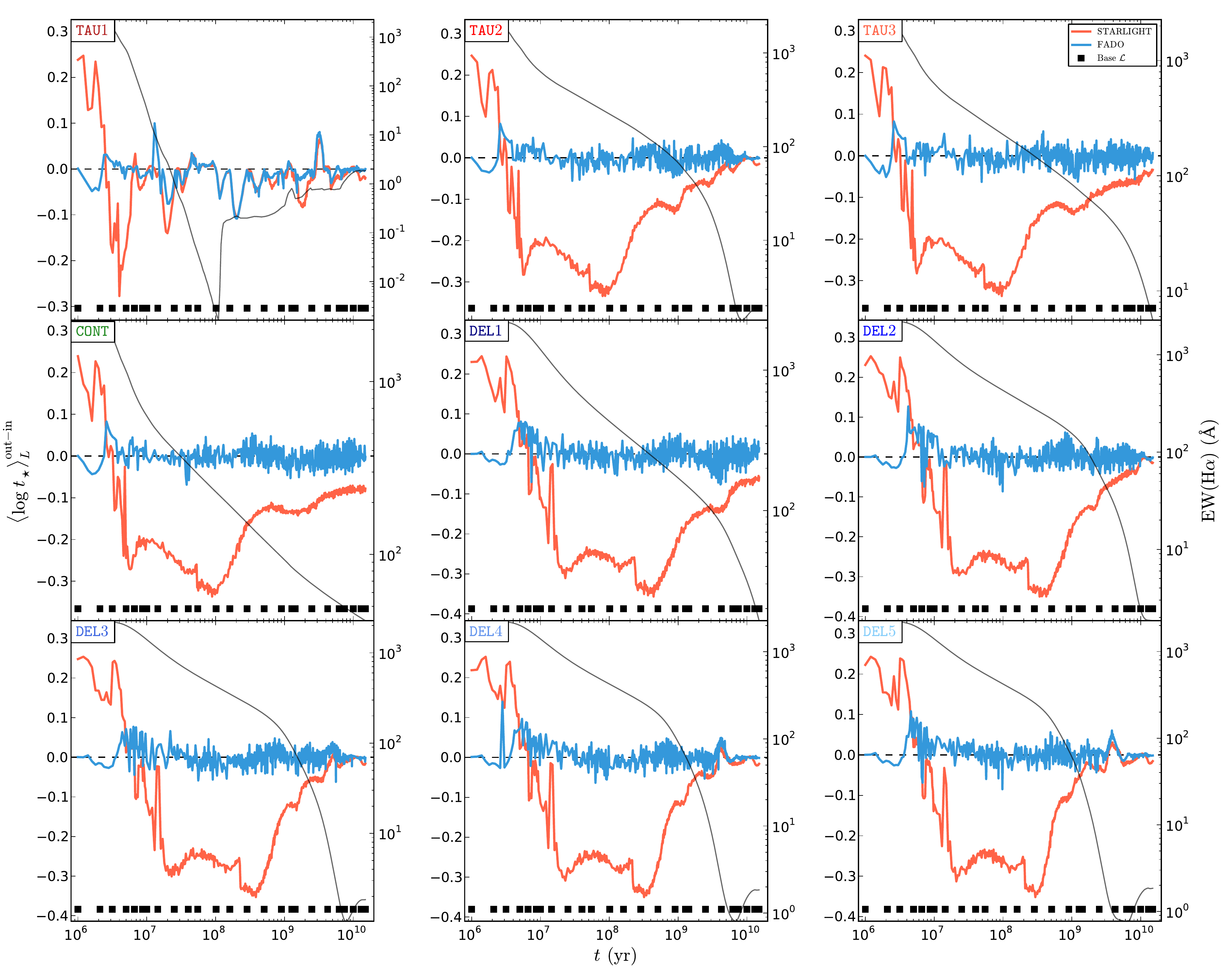}
\caption{Difference in the light-weighted mean stellar age $\langle\log \, t_{\star}\rangle_L$ between \Starlight/\Fado\ ({out}) and \Rebetiko\ ({in}) values as a function of the CSP age $t$ for different SFHs. Legend details are identical to those in in Figure \ref{Fig:SFHs_-_Mstar_diff_vs_age}. }
\label{Fig:SFHs_-_tL_diff_vs_age}
\end{center}
\end{figure*}

\begin{figure*}
\begin{center}
\includegraphics[width=0.99\textwidth]{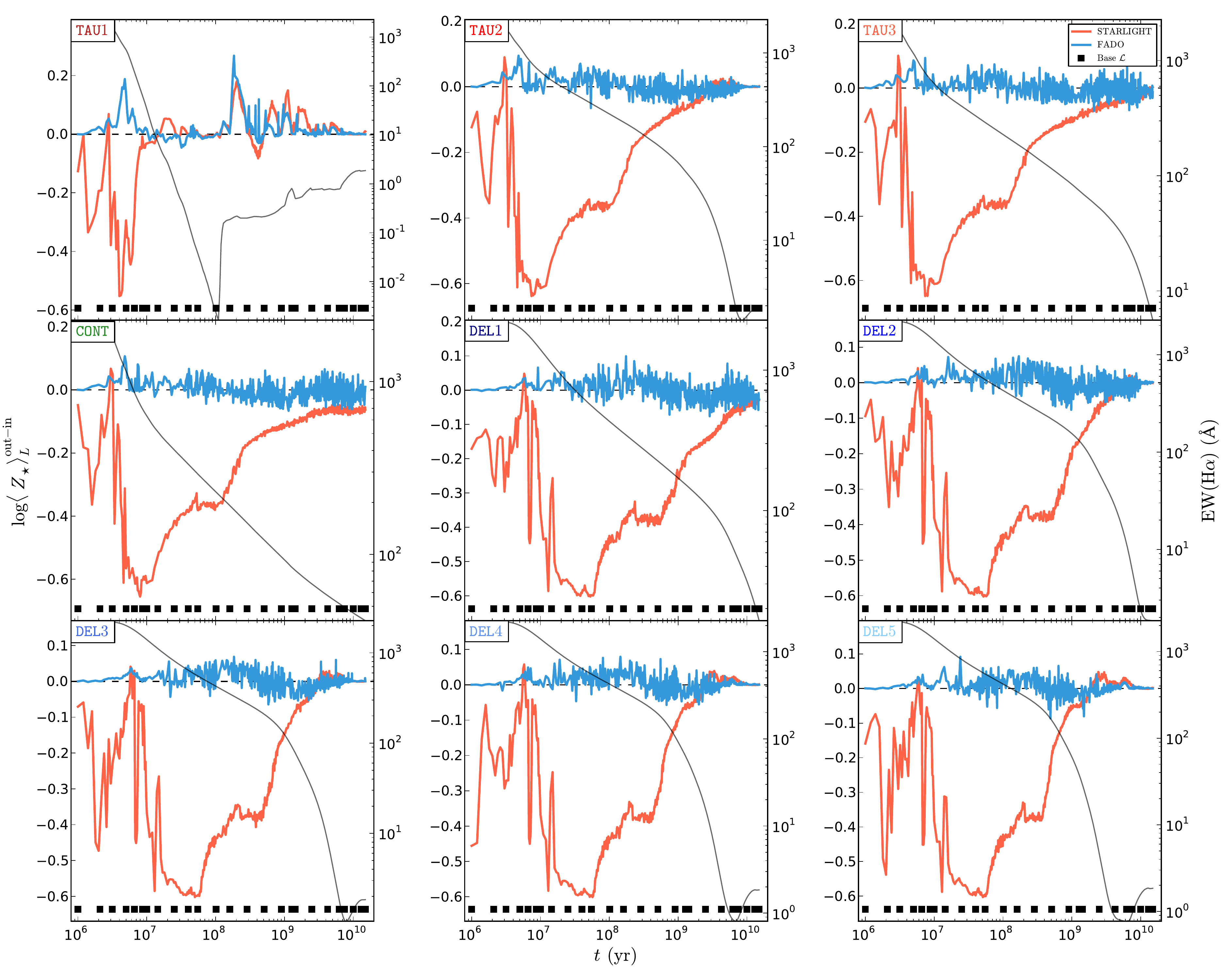}
\caption{Difference in the light-weighted mean stellar metallicity $\log \langle Z_{\star}\rangle_L$ between \Starlight/\Fado\ ({out}) and \Rebetiko\ ({in}) values as a function of the CSP age $t$ for different SFHs. Legend details are identical to those in in Figure \ref{Fig:SFHs_-_Mstar_diff_vs_age}. }
\label{Fig:SFHs_-_ZL_diff_vs_age}
\end{center}
\end{figure*}

\begin{figure*}
\begin{center}
\includegraphics[width=0.99\textwidth]{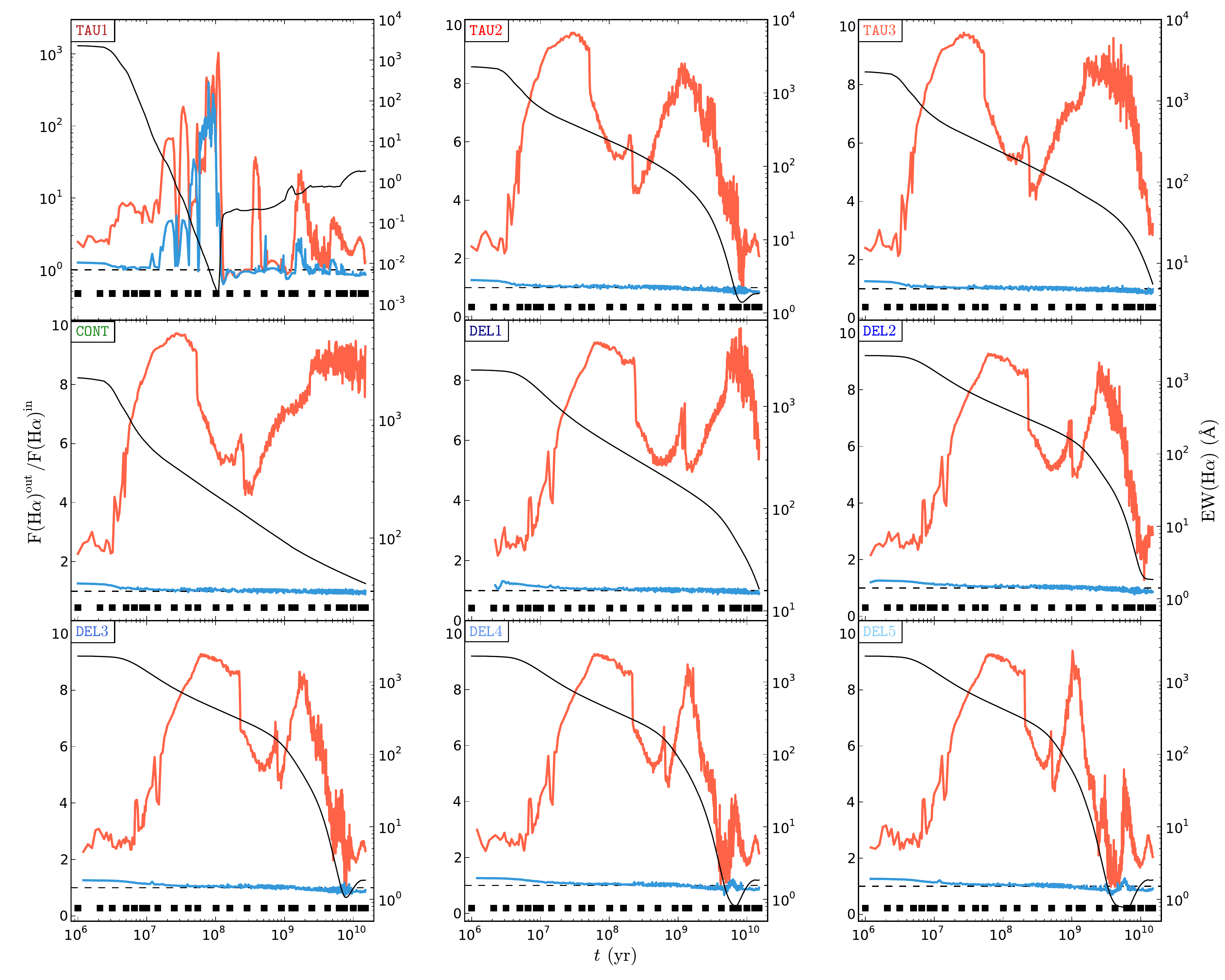}
\caption{Ratio of the \Starlight/\Fado\ (out) to \Rebetiko\ (in) H$\alpha$ fluxes F(H$\alpha$)  as a function of the CSP age $t$ for different SFHs. Legend details are identical to those in in Figure \ref{Fig:SFHs_-_Mstar_diff_vs_age}.}
\label{Fig:SFHs_-_FHalpha_vs_base_size}
\end{center}
\end{figure*}

	Another facet of the testuits focussed on the impact that the adopted parametric SFH in \Rebetiko\ has on the estimated physical and evolutionary properties of galaxies using \PS\ codes. Figures \ref{Fig:SFHs_-_Mstar_diff_vs_age}--\ref{Fig:SFHs_-_FHalpha_vs_base_size} show stellar mass, age and metallicity spectral fitting results and F(H$\alpha$) measurements from both \SL\ and \Fado\ using \baseL\ for the SFHs presented in \cite{Cardoso2017}. A comparison of these results with Figs. \ref{Fig:COVER_-_Mstar_diff_vs_age}--\ref{Fig:COVER_-_FHalpha_vs_base_size} shows that the SFH limiting cases of an instantaneous burst and continuous SFHs captures the variance within the stellar properties biases discussed in Secs. \ref{Sec:Results} \& \ref{Sec:Discussion}.

% FIGURE  B1 - - - - - - - - - - - - - - - - - - - - - - - - - - - - - - - - - - - - - - - - - - - - - - - - - - - - - - - - - - - - - - - - - - - - - - - - - - - - - - - 
% - - - - - - - - - - - - - - - - - - - - - - - - - - - - - - - - - - - - - - - - - - - - - - - - - - - - - - - - - - - - - - - - - - - - - - - - - - - - - - - - - - - - - - - - -
% FIGURE  B2 - - - - - - - - - - - - - - - - - - - - - - - - - - - - - - - - - - - - - - - - - - - - - - - - - - - - - - - - - - - - - - - - - - - - - - - - - - - - - - - 
% - - - - - - - - - - - - - - - - - - - - - - - - - - - - - - - - - - - - - - - - - - - - - - - - - - - - - - - - - - - - - - - - - - - - - - - - - - - - - - - - - - - - - - - - -
% FIGURE  B3 - - - - - - - - - - - - - - - - - - - - - - - - - - - - - - - - - - - - - - - - - - - - - - - - - - - - - - - - - - - - - - - - - - - - - - - - - - - - - - - 
% - - - - - - - - - - - - - - - - - - - - - - - - - - - - - - - - - - - - - - - - - - - - - - - - - - - - - - - - - - - - - - - - - - - - - - - - - - - - - - - - - - - - - - - - -
% FIGURE  B4 - - - - - - - - - - - - - - - - - - - - - - - - - - - - - - - - - - - - - - - - - - - - - - - - - - - - - - - - - - - - - - - - - - - - - - - - - - - - - - - 
% - - - - - - - - - - - - - - - - - - - - - - - - - - - - - - - - - - - - - - - - - - - - - - - - - - - - - - - - - - - - - - - - - - - - - - - - - - - - - - - - - - - - - - - - -
% FIGURE  B5 - - - - - - - - - - - - - - - - - - - - - - - - - - - - - - - - - - - - - - - - - - - - - - - - - - - - - - - - - - - - - - - - - - - - - - - - - - - - - - - 
% - - - - - - - - - - - - - - - - - - - - - - - - - - - - - - - - - - - - - - - - - - - - - - - - - - - - - - - - - - - - - - - - - - - - - - - - - - - - - - - - - - - - - - - - -
% FIGURE  B6 - - - - - - - - - - - - - - - - - - - - - - - - - - - - - - - - - - - - - - - - - - - - - - - - - - - - - - - - - - - - - - - - - - - - - - - - - - - - - - - 
% - - - - - - - - - - - - - - - - - - - - - - - - - - - - - - - - - - - - - - - - - - - - - - - - - - - - - - - - - - - - - - - - - - - - - - - - - - - - - - - - - - - - - - - - -

\end{document}